\documentclass[journal]{IEEEtran}

\usepackage{amsmath}
\usepackage{balance}
\usepackage{graphicx}
\usepackage{cite}
\usepackage[hyphens]{url}
\usepackage{hyperref}
\usepackage{gensymb}
\usepackage{textcomp}
\usepackage{color}
\usepackage[caption=false,font=footnotesize]{subfig}
\begin{document}
%
\title{Vehicular LTE Connectivity Analysis in Urban and Rural Environments using USRP Measurements}
%
%
%

\author{\IEEEauthorblockN{
        Karthik~Vasudeva\IEEEauthorrefmark{1}\IEEEauthorrefmark{2},
        Ozgur~Ozdemir\IEEEauthorrefmark{2},
        Sugan~R.~S.~Chandar\IEEEauthorrefmark{2},
        Fatih~Erden\IEEEauthorrefmark{2},
        Ismail~Guvenc\IEEEauthorrefmark{2}}
        
      \IEEEauthorblockA{\IEEEauthorrefmark{1}Electrical and Computer Engineering, Florida International University, Miami, FL, USA\\
      \IEEEauthorrefmark{2} Electrical and Computer Engineering, North Carolina State University, Raleigh, NC, USA}

}
\maketitle
\begin{abstract}
The intelligent transportation system (ITS) offers a wide range of applications related to traffic management, which often require high data rate and low latency. The ubiquitous coverage and advancements of the Long Term Evolution (LTE) technology have made it possible to achieve these requirements and to enable broadband applications for vehicular users. In this paper, we perform field trial measurements in various different commercial LTE networks  using software defined radios (SDRs) and report our findings. First, we provide a detailed tutorial overview on how to post-process SDR measurements for decoding broadcast channels and reference signal measurements from LTE networks.  We subsequently describe the details of our measurement campaigns in urban, sub-urban, and rural environments. Based on these measurements, we report joint distributions of base station density, cellular coverage, link strength, disconnected vehicle duration, and vehicle velocity in these environments, and compare the LTE coverage in different settings.  Our experimental results quantify the stronger coverage, shorter link distances, and shorter duration of disconnectivity in urban environments when compared to sub-urban and rural settings.  
\end{abstract}

\begin{IEEEkeywords}
3GPP, LTE connectivity, reference signal received power (RSRP), road-side unit (RSU), USRP.
\end{IEEEkeywords}

%
\IEEEpeerreviewmaketitle

\section{Introduction}
The rapid advancement in information and communication technologies stimulated a connectivity solution for vehicles to improve traffic safety and management. Such a solution is a key aspect of the intelligent transportation system (ITS), which is said to transform the way we travel today. The vehicles communicate not only with each other but also with nearby road-side unit (RSU) infrastructures to warn drivers against a potential accident or other unsafe events. In order to improve safety and mobility of the road travelers, there are various ongoing work on the connected vehicle pilot deployment programs in New York City, Tampa, and Wyoming~\cite{US_DOT_con_vehicle}. The US Department of Transportation~(USDOT) claims that the connected vehicle technology has the potential to eliminate up to 80\% of the non-impaired crashes~\cite{US_DOT_con_outcome}. 

Potential incident warnings should be instantly indicated to drivers in the rapidly varying vehicular channel. Therefore, reliability and timeliness are of utmost importance in vehicle-to-everything~(V2X) communication where wireless connective framework plays a crucial role. The US Federal Communications Commission~(FCC) allocated separate 75~MHz bandwidth in the 5.850-5.925~GHz band to implement ITS applications based on dedicated short-range communication~(DSRC) framework~\cite{FCC_DSRC}. The DSRC framework uses IEEE 802.11p, which is a modified version of the Wi-Fi standard and designed for wireless access in vehicular environments~(WAVE). It ensures interoperability between devices from different manufacturers and provides fast network establishment in the rapidly changing mobile environment~\cite{DSRC_USA}. In DSRC network, the onboard unit~(OBU) in a vehicle receives disseminated safety messages either from neighboring vehicles or the RSU infrastructure to determine emergency events, and the warning is sent to drivers in case of emergency. However, there are some concerns over its contention-based scheduling mechanism where it is difficult to achieve critical latency requirements in high-density traffic scenarios. Therefore, congestion controlling schemes need to be implemented in DSRC~\cite{weinfield2011adaptive}.

Alternatively, the cellular-based V2X communication~(C-V2X) started by the third generation partnership project (3GPP) is working on leveraging the long term evolution~(LTE) framework for vehicular communications~\cite{3GPP_CV2X}. As a result, the present deployed LTE infrastructure can be configured to support vehicular connectivity, thereby reducing the number of RSU installation. Further, large coverage scope and high capacity of the LTE make it a suitable fit to support low-latency safety-critical services for high-speed vehicles. Moreover, its higher penetration rate facilitates reliable packet delivery at road intersections, unlike DSRC technology, which suffers due to non-line-of-sight~(NLOS) conditions and shorter coverage. These advantages have sparked interest recently among leading automotive and cellular companies, e.g., Qualcomm and Ford, to deploy C-V2X in Colorado~\cite{qualcomm_V2X}.



The architecture enhancements to enable LTE-based V2X applications are given in~\cite{TS23285}. The vehicles may receive V2X messages either via unicast or broadcast setup called multimedia broadcast/multicast service (MBMS)~\cite{TS23246} over LTE-Uu interface for vehicle-to-infrastructure (V2I) communication. In addition, many additional modes have been introduced for direct vehicle-to-vehicle (V2V) communication in~\cite{TS36213}. In one of the modes, V2V resource allocation is managed by the base station (BS); thus the availability of coverage is crucial to ensure collision-free transmission which is unlikely in the case of the contention-based approach employed in DSRC technology. Therefore, ubiquitous coverage is desirable for efficient V2X communication.

Usually, the propagation losses and interference are the major limiting factors for vehicular communication, and their evaluation requires BS location information. Recently, BS location data are publicly available, e.g., in OpencellID~\cite{opencellid} and CellID finder~\cite{cellid_finder}, through estimation techniques. OpenCellID in particular incorporated crowd-sourced data collection scheme to estimate the location of the BSs and developed a database consisting of GPS coordinates of all the BS. The location estimates in the OpenCellID database are accessed using the global identity, which is periodically disseminated by the BS, and it is known as E-UTRAN Cell Global Identifier (ECGI) in LTE.

In this paper, we study the vehicular connectivity performance of the LTE networks based on the drive test measurements carried out using a universal software defined radio (USRP). The measurements are post-processed to decode identity information of the BSs and to evaluate their coverage strength. Primarily, we aim to decode unique identification of BSs specified by the ECGI and to extract their locations from OpenCellID. The results from the rural, semi-urban, and urban areas of the drive test campaign are obtained to investigate the coverage performance and its impact on the temporal connectivity of vehicular users. Our results show that V2I communication is better supported by LTE network deployments in semi-urban and urban compared to rural areas. 



The paper is organized as follows. Section~\ref{Sec: Lit_Rev} provides the literature review. Section~\ref{Sec: Meas_setup} describes the drive test measurement setup and the recording process. The post-processing steps performed in MATLAB to evaluate the BS information are given in Section~\ref{Sec:Post_process}. In Section~\ref{Sec: Res_Analy}, drive test results assessing the performance of the LTE infrastructure are presented, followed by concluding remarks in the last section.


\section{Literature Review}\label{Sec: Lit_Rev}

The vehicular connectivity performance of Wi-Fi networks for Internet access is studied in~\cite{Amdouni_ICT_2010,bychkovsky_mobicom_2006,mahajan_sigcomm_2007} using drive test measurements. In~\cite{Amdouni_ICT_2010,bychkovsky_mobicom_2006}, active scanning is analyzed, where moving users send probe requests to access points (APs) within their vicinity for the connection attempt. The distribution of APs discovered during active scan were examined in geographical areas with different population distribution in~\cite{Amdouni_ICT_2010}, while temporal connectivity information for different vehicular speeds is investigated in~\cite{bychkovsky_mobicom_2006}. Alternatively, connection requests can be sent by users in the coverage range of AP after successfully receiving the  beacon. The connection performance of this approach of passively scanning the APs is studied in ~\cite{mahajan_sigcomm_2007}. The intermittent connectivity was observed at several locations during the motion of the vehicle, and the possibility of poor link period is estimated using the knowledge of past connection measurements. As a result, the connection framework in Wi-Fi networks is suitable to support delay-tolerant applications like web browsing, messaging, etc., in vehicular scenarios.


On the other hand, the DSRC designed to support vehicular safety applications employ amended version of IEEE 802.11 (Wi-Fi) standard, called IEEE 802.11p, to address the safety-critical communication requirements mentioned in~\cite{Araniti_CommMag_2013,Zheng_Surv_2015}. The field test campaigns were performed in IEEE 802.11p-based vehicular network to examine the impact of vehicular speed on latency and throughput of the V2I communication, respectively in~\cite{Lin_ICC_2010,Sukuvaara_ICUFN_2012}. The reported results confirm that normal range vehicular speeds do not have a direct influence on the performance of V2I communication. The work in~\cite{Gozalvez_CommMag_2012,Shivaldova_VehMag_2013,Sassi_IWCMC_2015,Shivaldova_PIMRC_2012,Shivaldova_ICC_2015,Sepulcre_VNC_2013} evaluate packet delivery ratio (PDR) to determine the connectivity performance of V2I communication, and it is studied in various test environment and system settings of IEEE 802.11p standard.

Initially, the results in~\cite{Gozalvez_CommMag_2012} show that the NLOS conditions characterized by obstacles like trees, heavy traffic conditions, etc., degrade the connectivity range. The height of the antenna and changes in terrain elevation are also known to affect the visibility between vehicles and RSU. Usually, the vehicular channels are highly time varying and as a result, increasing either the throughput or packet size settings of the RSU have a negative influence on the link connectivity as reported in~\cite{Shivaldova_VehMag_2013,Sassi_IWCMC_2015}. Further, the results in~\cite{Shivaldova_PIMRC_2012} show that high-beam antennas improve the coverage range; however, its precise positioning, which can be easily influenced by environmental conditions, becomes more critical. Lastly, the connectivity performance is estimated using the Gilbert model in~\cite{Shivaldova_VehMag_2013,Shivaldova_ICC_2015}, while artificial neural and Bayesian networks are employed in~\cite{Sepulcre_VNC_2013} to estimate the connectivity range considering different context information from drive test measurements for training the networks. 

The DSRC technology is usually suitable for short transmission ad hoc type communication because serving users with V2X applications in a large coverage area requires a centralized approach of allocating radio resources for handling the channel congestion, which is a problem in DSRC. Therefore, the DSRC is not ideal for use in V2I applications for long-range communications. Alternatively, the LTE-V2X standard is capable of catering reliable safety services to many vehicular users due to its wider coverage and high capacity. In~\cite{Hu_IMCEC_2018,Cecchini_VNC_2017}, V2V communication performance is evaluated using packet reception rate (PRR), and the simulation results show that V2V in coverage perform better compared to out of coverage where users autonomously select their radio resources without the assistance from BS. Further, drive test experiment results in~\cite{Shi_CAC_2017} confirm that the PDR performance of LTE cellular-based vehicular services is better than DSRC in V2I connectivity situation at the road intersections. The LTE-V2X standard is still in trial stages, and currently, there are no real-world deployments which offer vehicular safety services. Therefore, studying vehicular connectivity with the existing LTE deployments will help to determine the important factors influencing the vehicular communication performance.

An experimental testbed is developed in~\cite{xu2017dsrc_lte,liu_sage_2016} to obtain the link statistics of DSRC and LTE frameworks in vehicular safety applications. The results show that the link performance of LTE is better compared to DSRC even in NLOS conditions, at the expense of compromise in latency. The measurement campaigns were performed in~\cite{Lauridsen_CommMag_2017,Akselrod_VTC_2017,Berisha_LTE_coverage,Schaffner_COMCAS_2015} to study various performance indicators of LTE networks such as latency, throughput, and coverage in urban and rural areas. In~\cite{Lauridsen_CommMag_2017}, the core and the access network delays are evaluated, and the results reveal that the network enhancements are required to obtain latency performance, which can support vehicular safety use cases. The download speed is analyzed in~\cite{Akselrod_VTC_2017}, and results show that its variability with distance is more significant in urban compared to rural areas. The results also indicate that parameters like signal quality and bandwidth influence the throughput. Further, the maximum coupling loss between the user and the base station determined in~\cite{Lauridsen_CommMag_2017} show that LTE provides 99\% coverage to outdoor and road users. The signal strength results in~\cite{Berisha_LTE_coverage,Schaffner_COMCAS_2015} indicate higher penetration loss in sub-urban/rural compared to urban areas since channel propagation characteristics differ in these radio environments.

The pathloss characteristics of V2I channel were evaluated in~\cite{Yangi_URSIGASS_2017,Rubio_propagation_2015,lee_jkosme_2016} (see also \cite{anjinappa2018millimeter,anjinappa2018angular} for millimeter wave vehicular channels). It is shown that the path loss is affected by the antenna height, and its behavior can be described using the two-ray ground reflection model. In rural/suburban areas, multipath components are not prominent compared to urban areas as depicted by the average power delay profile results in~\cite{Yangi_URSIGASS_2017}. This is because of the lesser obstacles in the flat terrain of rural/suburban areas, and scattering mostly occurs in the close area around the mobile user~\cite{Mecklenbrauker_IEEE_proc_2011}. Further, in the NLOS scenario obstruction from neighboring vehicles case, measurements in~\cite{Aygun_VTC_2016} are quite in agreement with the knife-edge model in~\cite{ITU_diffraction}, which accounts for additional attenuation produced by diffraction. Simulations are performed in~\cite{Mecklenbrauker_IEEE_proc_2011} to evaluate bit error rate (BER) performance in NLOS for V2I channel. The results show that the channel estimation performance of the IEEE 802.11p pilot pattern is degraded, and BER does not improve with increasing SNR.

\section{Drive Test Measurement Setup and Data Collection Process} \label{Sec: Meas_setup}
\begin{figure}[t]
  \centering
  \includegraphics[width=3.45in]{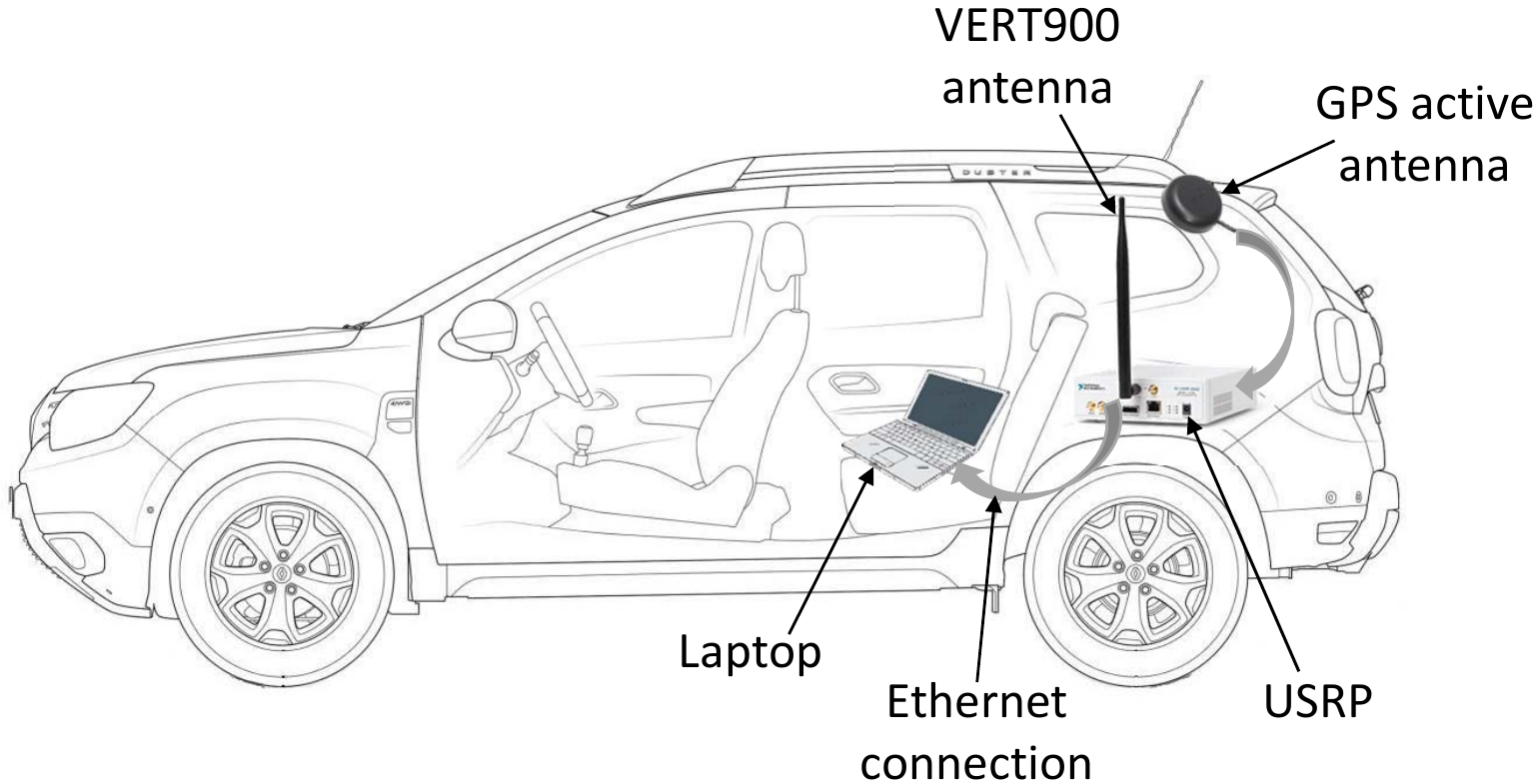}
  \caption{Drive test measurement setup with USRPs.}
  \label{fig:drive_test_setup}
\vspace{-3mm}
\end{figure}

The drive test measurements are obtained using a USRP N210 with an integrated GPS disciplined oscillator (GPSDO) module, which offers a flexible programmable platform to implement custom functions using onboard FPGA~\cite{Ettus_USRP,Pidanic_2017,nadisanka_lte,kumar2014lte}. The GPSDO module is connected to a 3V active antenna for faster acquisition of location information. The measurement setup is shown in Fig.~\ref{fig:drive_test_setup}. The active antenna improves the signal reception performance and thereby yielding faster locking on to GPS~satellites. The measurements are collected using a laptop through a Gigabit Ethernet connection, and both the USRP and the laptop are powered through an uninterruptible power supply (UPS) during the drive test.

\begin{figure}[t]
  \centering
  \includegraphics[width=2.7in]{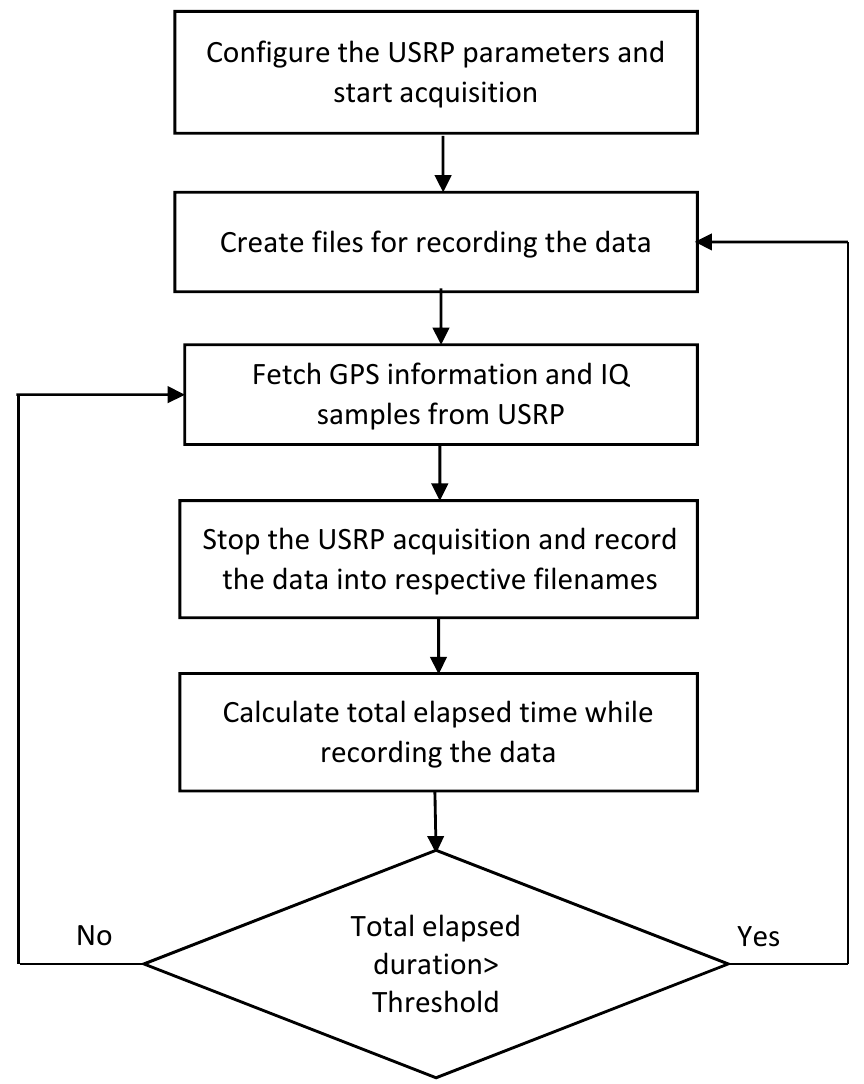}
  \caption{Measurement collection process in SDR experiments.}
  \label{fig:USRP_fetch_process}
\end{figure}

\begin{figure}[t]
  \centering
  \captionsetup{justification=centering,margin=2cm,labelfont=sc}
  \includegraphics[width=3.2in]{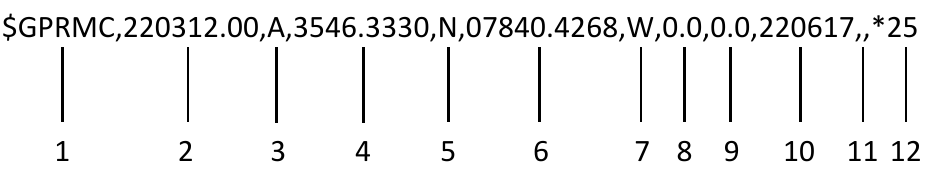}
  \caption{NMEA RMC sentence format.}
   \vspace{-4mm}
  \label{fig:GPS_RMC}
\end{figure}

The measurement collection is performed in LabVIEW running on the laptop, and the process is shown in the flowchart in Fig.~\ref{fig:USRP_fetch_process}. The GPS and IQ data samples from the USRP are continuously recorded in respective files in CSV format, while total elapsed time is monitored. If the elapsed duration exceeds a threshold, then new files are created, and recordings are carried over to the new files. This avoids the difficulty in handling large files during post-processing of the measurement data. The GPS information is obtained using the USRP driver, which returns the standard National Marine Electronics Association (NMEA) sentences from the GPSDO module. Fig.~\ref{fig:GPS_RMC} shows the generated recommended minimum data (RMC) format type where each data segment is separated by a comma. The segments are described in Table~\ref{Tab:RMC_description}. The latitude and longitude field values are converted to decimal format, and sign convention is applied as indicated by the N/S, E/W data fields. The latitude and longitude values are considered negative for south and west, respectively, which are then recorded along with the speed value.

\begin{table}[t]
\caption{RMC sentence description (see Fig.~\ref{fig:GPS_RMC}).}
 \centering
\begin{tabular}{|c| l|}
 \hline
  {\bfseries Field \#} & {\bfseries Description}\\
 \hline
  $1$ & Protocol header type \\ \hline $2$ & UTC time format \\ \hline
 $3$ & Status (A=valid;V=invalid) \\ \hline $4$ & Latitude (DD\degree MM.MMMM') \\ \hline
 $5$ & North/South (N/S) indicator \\ \hline $6$ & Longitude (DD\degree MM.MMMM')\\ \hline
 $7$ & East/West (E/W) indicator \\ \hline $8$ & Speed in knots\\ \hline
 $9$ & Course over ground \\ \hline $10$ & Date (ddmmyy) \\ \hline
 $11$ & Magnetic variation \\ \hline $12$ & Checksum \\ \hline
\end{tabular}
 \vspace{-2mm}
\label{Tab:RMC_description}
\end{table}



Further, the USRP parameters, like sampling rate, carrier frequency, and number of samples per fetch, are initially configured before starting the I/Q signal acquisition. It is crucial to set the ideal parameters to accomplish successful decoding of primary system information and ECGI, which are specified in the master information block (MIB) and system information block type 1 (SIB1), respectively. The BS broadcasts this information periodically, and it is important to study their scheduling information in time and frequency. First, we study the scheduling pattern of MIB and SIB1 in frequency to determine the sampling rate of the USRP. The MIB is carried by the physical broadcast channel (PBCH) transmitted in central six OFDM resource blocks (RBs), and therefore corresponding sampling rate of $1.92$~MHz is required, regardless of the system bandwidth at the BS. The USRP N210 does not support this sampling rate since it allows only integer decimation factor $N$ of the master clock rate given by $\frac{100e^{6}}{N}$. Thus, the nearest sampling rate of $2$~MHz is set during the measurement recording in a drive test. In contrast, SIB1 is transmitted using physical downlink shared channel (PDSCH) over the entire system bandwidth indicated by the MIB. After a few preliminary drive tests, we observe base station's bandwidth to be $10$~MHz. Therefore, we set the sampling rate to $16.67$~MHz.




Next, we determine the total capture duration using the standard transmission time pattern of MIB and SIB1 in LTE for their successful extraction. The MIB is transmitted every $40$~ms, and within this period, its repetitions are transmitted over each radio frame to allow soft combining for decoding the PBCH in low signal strength conditions. The soft combining of PBCH is performed taking four sequential radio frames and advancing this window until successful decoding is attained~\cite{LTEBook}. As a result, we fix the capture duration of the USRP to $80$~ms. A similar approach is followed for SIB1 extraction, where the USRP is configured to record $160$~ms of I/Q signal data since it is transmitted with a periodicity of $80$~ms. Finally, two sets of I/Q signal measurements are recorded for extracting MIB and SIB1. The I/Q data for MIB extraction contain $80$~ms sampled at $2$~MHz, whereas for SIB1 contain $160$~ms sampled at $16.67$~MHz. In the next section, processing steps to decode MIB and SIB1 from the corresponding measurements are provided.

\vspace{-1mm}
\section{Post Processing of Drive Test Measurements}\label{Sec:Post_process}
The primary goal of the post processing is to unambiguously identify the BSs and analyze vehicular communication performance from the measurement data. The ECGI information enables unambiguous identification, and it is extracted from the measurements performed for decoding the SIB1 as mentioned in Section~\ref{Sec: Meas_setup}. Further, the vehicular communication performance is determined based on the achievable coverage from the measurement data by decoding the MIB as explained in the following section.

\vspace{-4mm}
\subsection{Decoding the MIB}\label{Subsec:PBCH_dec}
\begin{figure}[t]
  \centering
  \includegraphics[width=3.48in]{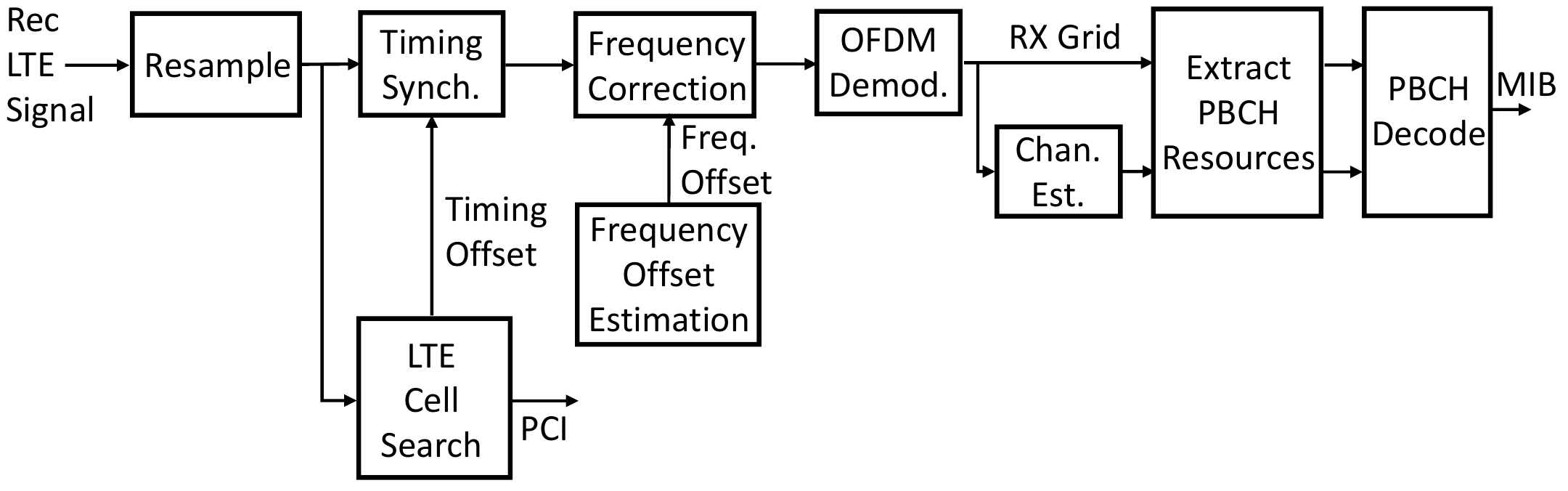}
  \caption{MIB decoding flowchart using the recorded LTE I/Q data from SDRs.}
  \label{fig:MIB_decode_model}
\end{figure}

The primary system information required for the initial access to the cell is specified in the MIB. Therefore, successful decoding of the MIB is checked to determine BS coverage from the measurement data. The standard LTE procedure is followed for MIB decoding using the MATLAB LTE toolbox~\cite{MatlabUSRP} as shown in Fig.~\ref{fig:MIB_decode_model}.

\begin{figure}[t]
\centering
\includegraphics[width=3.2in]{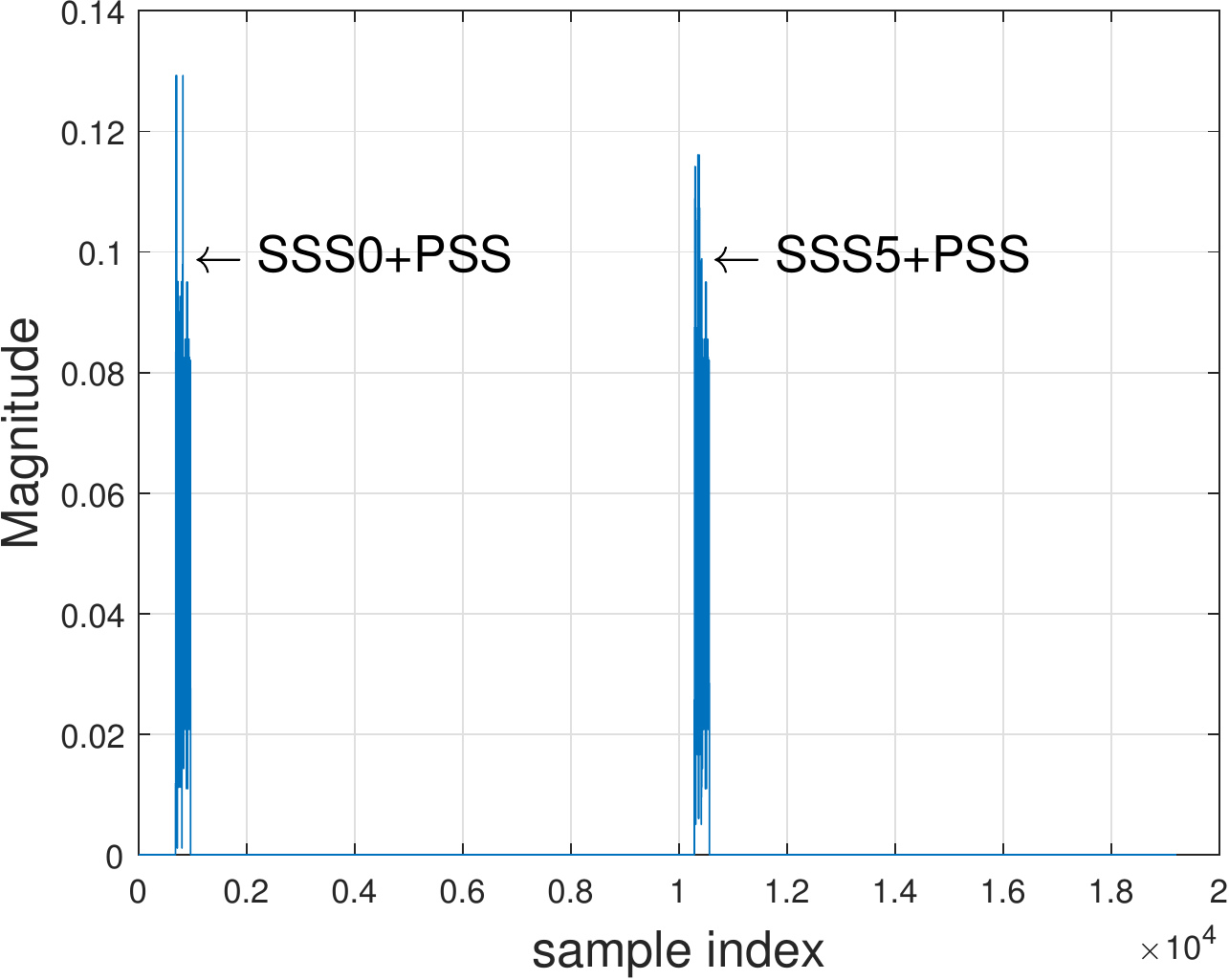}
\caption{Correlation results obtained for a particular PSS and SSS in a frame length of 19200 samples.}
\label{fig:fig4}
\vspace{-3mm}
\end{figure}

Initially, the recorded measurements dedicated to MIB extraction are resampled to $1.92$~MHz due to sampling rate restrictions of the USRP, as explained in Section~\ref{Sec: Meas_setup}. Then the cell search procedure is performed to evaluate the physical cell identity (PCI) and timing offset. The PCI in LTE is determined using a combination of Primary Synchronization Signal (PSS) and Secondary Synchronization Signal (SSS). There are three variations of PSS signals, and depending on the variation, there are 168 SSS signals resulting in 504 unique combinations. The signal data are correlated with all the 504 combinations of PSS and SSS, and the one which yields the peak in correlation gives the PCI. Unlike PSS, the SSS signals transmitted in subframe 0 and 5 differ in the cyclic shifts. This enables to determine the timing offset of the LTE frame in the receiver. For instance, the correlation peaks for a particular PSS and SSS combination occurring in subframe 0 and 5 of a frame length equal to 19200 samples is shown in Fig.~\ref{fig:fig4}. We can observe that the highest correlation peak is achieved within subframe 0, and the corresponding peak location gives the direct estimate of timing offset. Otherwise, the peak location would have delayed to the subframe 5 of the next frame for calculating the timing offset.

Further, the frequency offset is performed to compensate for oscillator errors and doppler shift in the vehicular scenarios. It is evaluated based on correlating the cyclic prefixes. Suppose $x(n)$ is the resampled signal and the sample at $n=n_1$ is part of the CP, then ignoring channel and AWGN,

\begin{equation}
    x(n_1+128)=x(n_1)\mathrm{e}^{j\phi 128}
\end{equation}
where $\phi$ is the angle created in one sample due to the frequency offset between the transmitter and the receiver. Note that,
\begin{equation}
  -\arg(x(n_1)x^*(n_1+128))=128 \phi
  \label{phioffset:eq}
\end{equation}
and frequency offset is related to $\phi$ by,
\begin{equation}
\label{foffset:eq}
     \hat{f}_{\mathrm{offset}}=\frac{\phi}{2 \pi T_s}
\end{equation}
where $T_s=\frac{1}{1.92\times 10^6}$ seconds.



\begin{figure}[t]
\centering
\includegraphics[width=3.2in]{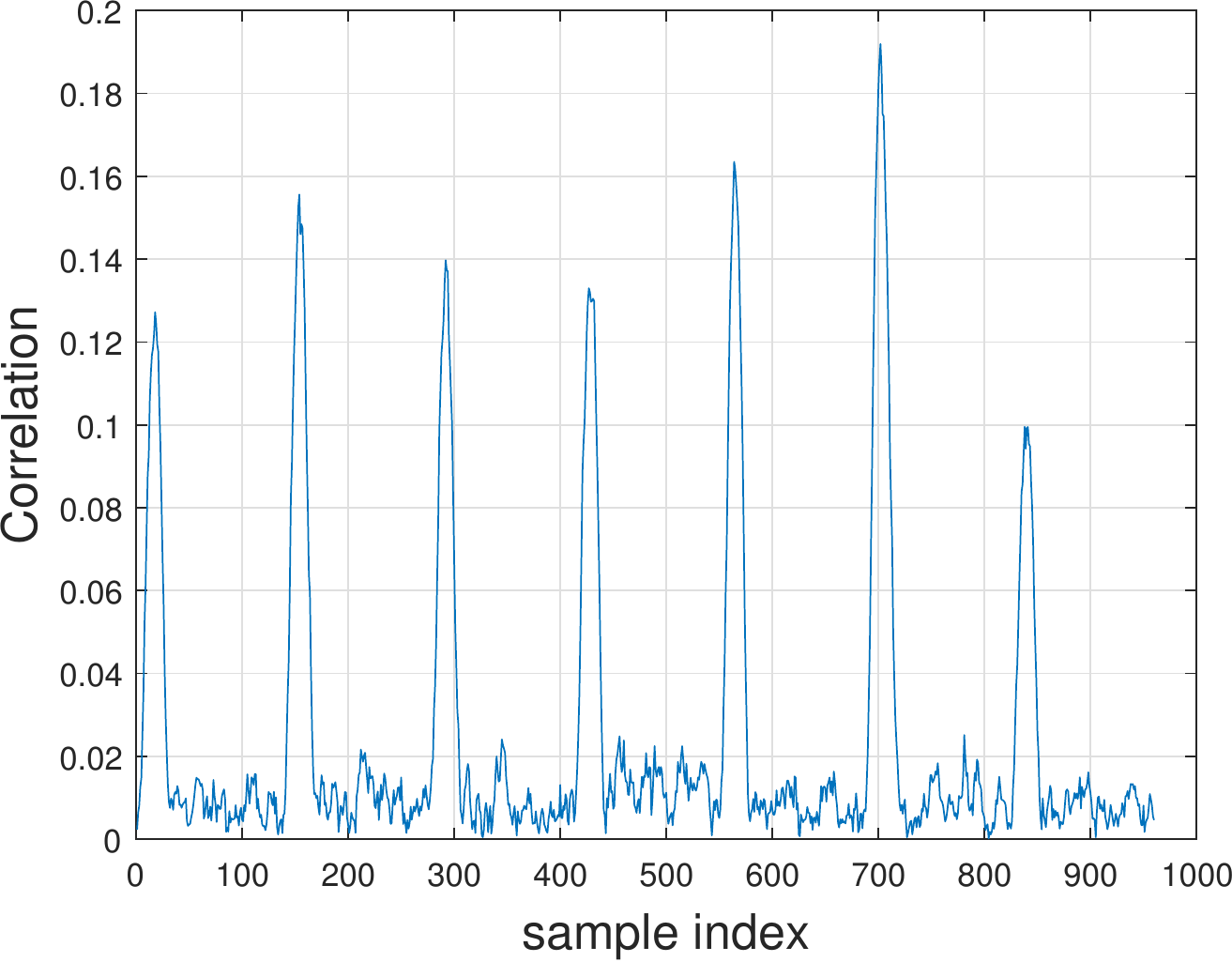}
\caption{Peak locations coincide with the CP location occurring in a half frame of 9600 samples.}
\label{fig:fig3}
\vspace{-3mm}
\end{figure}
For instance, Fig.~\ref{fig:fig3} shows the plot of $y(n)=x(n)x^*(n+128)$ integrated over all CPs in a slot and then averaged over all available slots in the received data whose length is equal to 960 samples. It clearly depicts the peaks of all CPs corresponding to seven OFDM symbols in a slot. The frequency synchronization is achieved by performing $x(n)\mathrm{e}^{j 2 \pi \hat{f}_{\mathrm{offset}} n T_s}$. In this approach, frequency synchronization is possible when $| 128\phi | \leq \pi \implies | \phi | \leq \frac{\pi}{128} $. Therefore maximum frequency offset is expressed as
\begin{equation}
     \hat{f}_{\mathrm{offset}} \leq \frac{\pi / 128 }{2 \pi T_s}=\frac{1.92\times 10^6}{2 \times 128}=7.5~\mathrm{kHz}~.
\end{equation}

Next, OFDM demodulation is carried out by removing CP and performing 128 point IFFT. The unused subcarriers are discarded resulting in a grid consisting of 72 subcarrier rows and columns indicating the signal duration. The channel estimates are then obtained for the resource grid to compensate for the imperfections in the channel. Further the PBCH extraction procedure is carried out, where initially PBCH symbols are extracted from the resource grid and subsequently inverse of PBCH processing is performed. It consists of deprecoding, symbol demodulation, and descrambling~\cite{MatlabPBCH_decode}. Lastly, MIB decoding is performed to obtain 24 MIB bits, out of which 10 bits are spare, and the remaining 14 bits actually signify three essential system information, namely
\begin{itemize}
    \item System bandwidth
    \item System Frame Number (SFN)
    \item Physical Hybrid Automatic Repeat Request Indicator Channel (PHICH) Configuration.
\end{itemize}
The system bandwidth information is carried in first 3 bits and its decimal equivalent, 0 to 5, corresponds to the RB set $N_{\rm RB}=~\{{6,15,25,50,75,100}\}$. The forth bit conveys whether the duration of PHICH is normal or extended, while the fifth and sixth bits indicate its multiplexing configuration, and the last  eight bits are used to determine the SFN. Apart from the system information, the transmit antenna number data are scrambled in CRC using an antenna specific mask~\cite{TS36212}. In order to obtain its data, CRC check is attempted for the standard values equal to 1, 2 and 4, and the corresponding true value is returned when the decoding is successful~\cite{MatlabPBCH_decode}.

Next, the BS coverage strength given by the reference signal received power (RSRP) is evaluated from the resource elements (REs) in the grid that carry cell-specific reference signals~\cite{LTEBook}. It is relatively measured in dB scale and all the results were obtained by keeping the same USRP gain setting. The V2I connectivity performance of the LTE is studied from the results explained in Section~\ref{Sec: Res_Analy}. However, unique identification of BSs is necessary to determine the impact of actual network deployments on the V2I connectivity performance. It is achieved by decoding ECGI information from the measurement data.

\vspace{-2mm}
\subsection{Decoding the PDSCH for ECGI extraction}\label{Subsec:PDSCH_dec}
The range of cell identification using the PCI is limited to 504, and it might be reused to identify the small cells in dense deployments scenario. This creates ambiguity and therefore BS broadcast ECGI contained within the system information indicated in the SIB1 for unique identification. The SIB1 is transmitted on PDSCH, and the decoding procedure given in~\cite{MatlabUSRP} is performed for its extraction from the recorded set of measurements specified in Section~\ref{Sec: Meas_setup}.

\begin{table}[t]
\caption{Sampling rates for different bandwidths in LTE.}
\vspace{-2mm}
\begin{center}
\begin{tabular}{|c| c| c|}
 \hline
  {\bfseries LTE Bandwidth} & {\bfseries LTE Sampling rate} & {\bfseries Number of}\\
 {\bfseries (MHz)} & {\bfseries (MS/s)} & {\bfseries Resource Blocks ($N_{\rm RB}$)}\\
 \hline
  $1.4$ & $1.92$ & $6$ \\
 \hline
 $3$ & $3.84$ & $15$ \\
 \hline
 $5$ & $7.68$ & $25$ \\
 \hline
 $10$ & $15.36$ & $50$ \\
 \hline
 $15$ & $23.04$ & $75$ \\
 \hline
 $20$ & $30.72$ & $100$ \\
 \hline
\end{tabular}
\end{center}
\label{Tab:USRP_samp_rate}
\end{table}

\begin{figure}[t]

\centering
\includegraphics[width=3.3in]{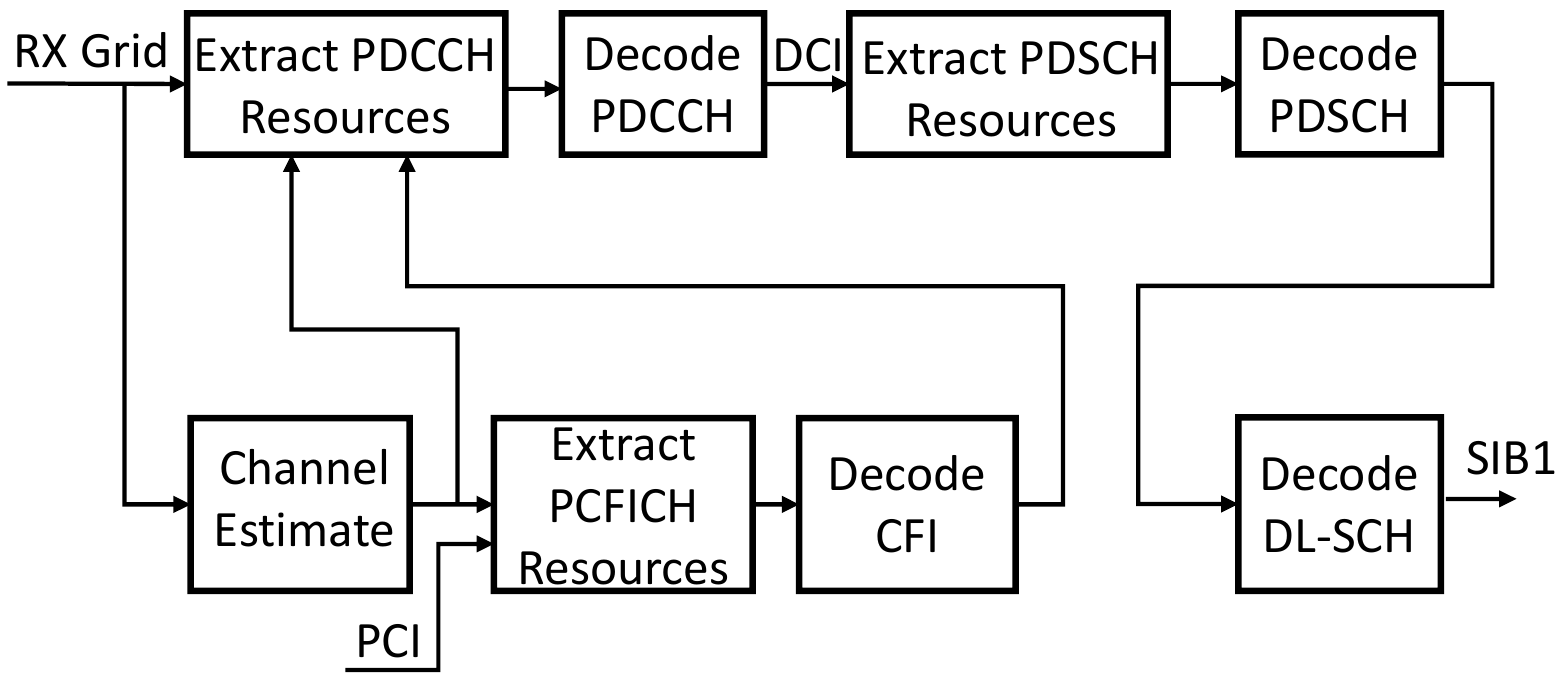}
\caption{Decoding of SIB1 information from SDR measurements.}
\label{fig:SIB1_decode}
\vspace{-3mm}
\end{figure}

Initially, the measurements are resampled according to the system bandwidth given in Table~\ref{Tab:USRP_samp_rate}, and the resource grid is obtained by repeating a sequence of processing steps for OFDM demodulation as specified in Section~\ref{Subsec:PBCH_dec}. The overview of PDSCH decoding procedure to acquire SIB1 is shown in Fig.~\ref{fig:SIB1_decode}. First, we extract the Physical Control Format Indicator Channel (PCFICH) REs, which are usually mapped into groups called as Resource Element Groups (REGs)~\cite{LTEBook}. There are four such groups, and each group contains four REs distributed across the frequency domain and their location indices are pre-defined with an offset related to PCI. Therefore prior knowledge of PCI is required for PCFICH symbols extraction. Next, inverse of PCFICH processing is performed, which includes deprecoding, symbol demodulation, and descrambling to obtain codeword sequences. Lastly, the codewords are decoded to determine the CFI value, and it usually has three standard values, i.e., 1, 2 or 3, which indicates the number of OFDM symbols transmitted in the physical downlink control channel (PDCCH)~\cite{LTEBook}.

Next, we acquire downlink control information (DCI) which is carried on the PDCCH to determine the scheduling information of the data transmitted on PDSCH. The DCI REs in the resource grid is mapped on to Control Channel Elements (CCEs) and their possible locations are checked in search space which comprises of user specific and common as defined in LTE~\cite{LTEBook}. Typically, the common search space carries DCI messages specifying the resource assignments of broadcast system information on the PDSCH. We perform blind decoding to check for all possible standard DCI formats using a broadcast identity called as System Information Radio Network Temporary Identifier (SI-RNTI)~\cite{LTEBook}. In order to perform blind decoding, we initially extract all the possible PDCCH resource elements from the resource grid across the entire subcarrier rows and the time span specified by the CFI~\cite{MatlabPDCCH_blindsearch}. Next, the inverse of PDCCH processing is performed consisting of deinterleaving, cyclic shifting, deprecoding, symbol demodulation and descrambling to obtain codeword sequences. Finally, blind decoding is performed to find all the possible DCI message format across each PDCCH candidate in common search space. If the CRC of the successful decoded PDCCH bits matches with the SI-RNTI identity then the corresponding returned DCI message indicates the resource assignments along with necessary decoding information for~SIB1.


Further, in order to decode the SIB1 message we first extract the PDSCH REs where their locations are provided by the DCI. Then decoding procedure is carried out by performing PDSCH channel inverse processing to obtain codeword sequences. It includes inverting the channel precoding, layer demapping and codeword separation, soft demodulation, and descrambling operations~\cite{MatlabPDSCH_decode}. Lastly, the codewords undergo decoding procedure consisting of rate recovery, turbo decoding, block concatenation, and CRC calculations~\cite{MatlabDLSCH_decode} to obtain SIB1 bits. In addition, decoding status is gathered for each codeword to check for error indicated by CRC check. This helps in error correction of incorrect codewords through soft combining with other repeated SIB1 messages captured in the other radio frames.

\begin{figure}[t]
 \centering
 \includegraphics[width=.43\textwidth]{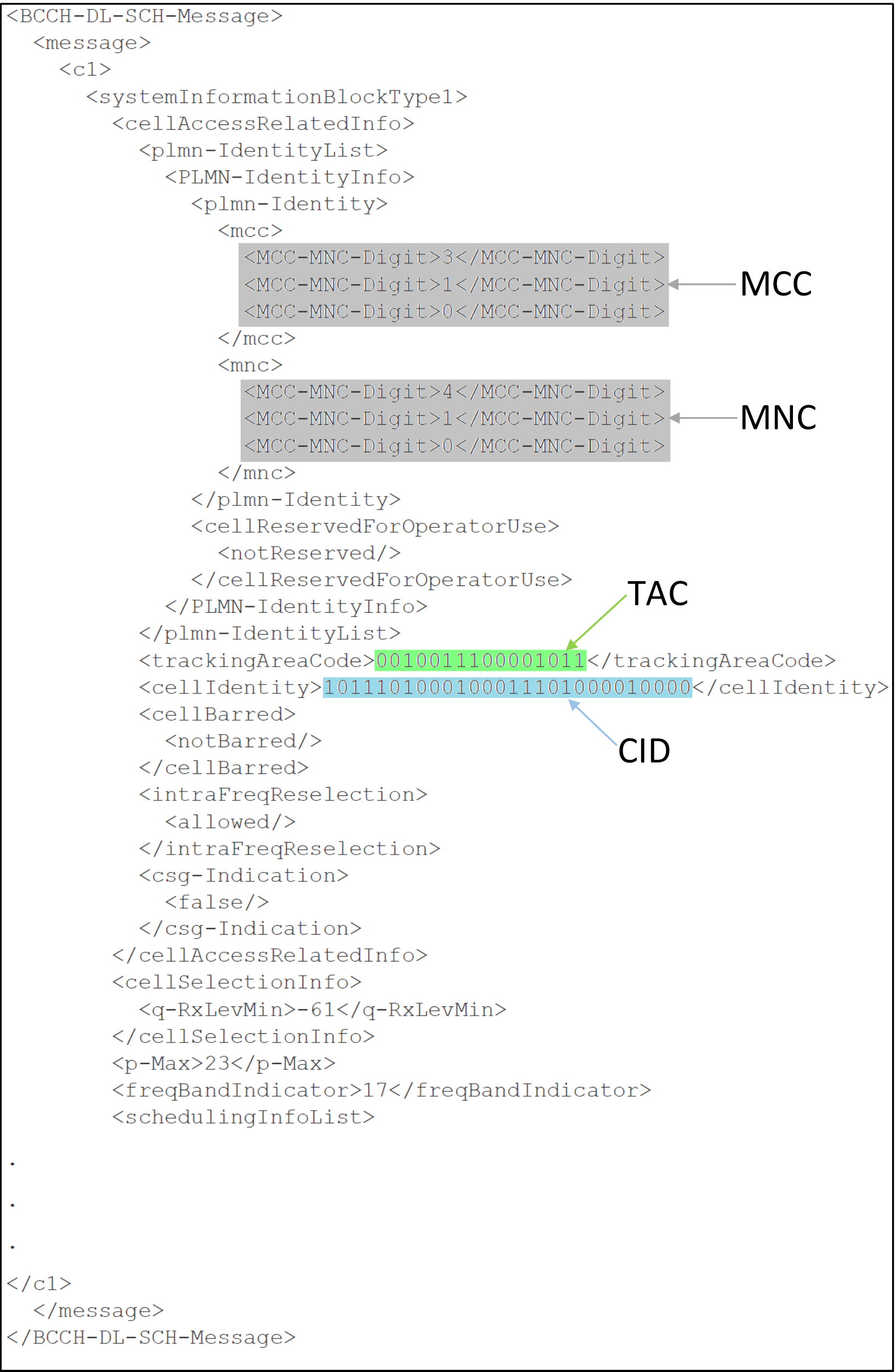}
 \vspace{-2mm}
 \caption{Decoded BCCH-DL-SCH message from SIB1 bits.}
 \label{fig:DL-SCH_message}
 \vspace{-3mm}
\end{figure}

Finally, we extract ECGI information by parsing SIB1 bits using ASN.1 compiler. The compiler is installed on linux platform by following the steps given in~\cite{peng_asn1_blog}. Initially, the SIB1 bits are stored in a file with .per format and decoding is performed to obtain BCCH-DL-SCH-Message as shown in Fig.~\ref{fig:DL-SCH_message}. The ECGI is given by Public Land Mobile Network (PLMN) information consisting of mobile country code (MCC), mobile network code (MNC), tracking area code (TAC) and cell identity (CID) to identify the cell within the PLMN~\cite{LTEBook}.

Using the ECGI information, we obtain BS location from the OpenCellID database and CellID finder~\cite{cellid_finder} and it is important to note that locations provided by database are approximate. Then we incorporate BS location results to analyze the effect of LTE network deployments on the performance of V2I communication shown in the following section.

\section{Results and Analysis}
\label{Sec: Res_Analy}

The drive test results are provided to investigate the impact of parameters such as frequency and mobility on V2I link performance in a small scale shown in Section~\ref{Subsec:Small_area_analysis}. Further, we present results of the large scale drive test measurements to study how LTE network deployments in rural, semi-urban and urban counties influence the V2I link performance explained in Section~\ref{Subsec:Large_area_analysis}.

\vspace{-2mm}
\subsection{Impact of LTE  Frequencies on V2I Link Performance} \label{Subsec:Small_area_analysis}

\begin{figure}[t]
\centering
\subfloat[{RSRP in the drive test scenario around NCSU campus.}]{\includegraphics[width=3.05in]{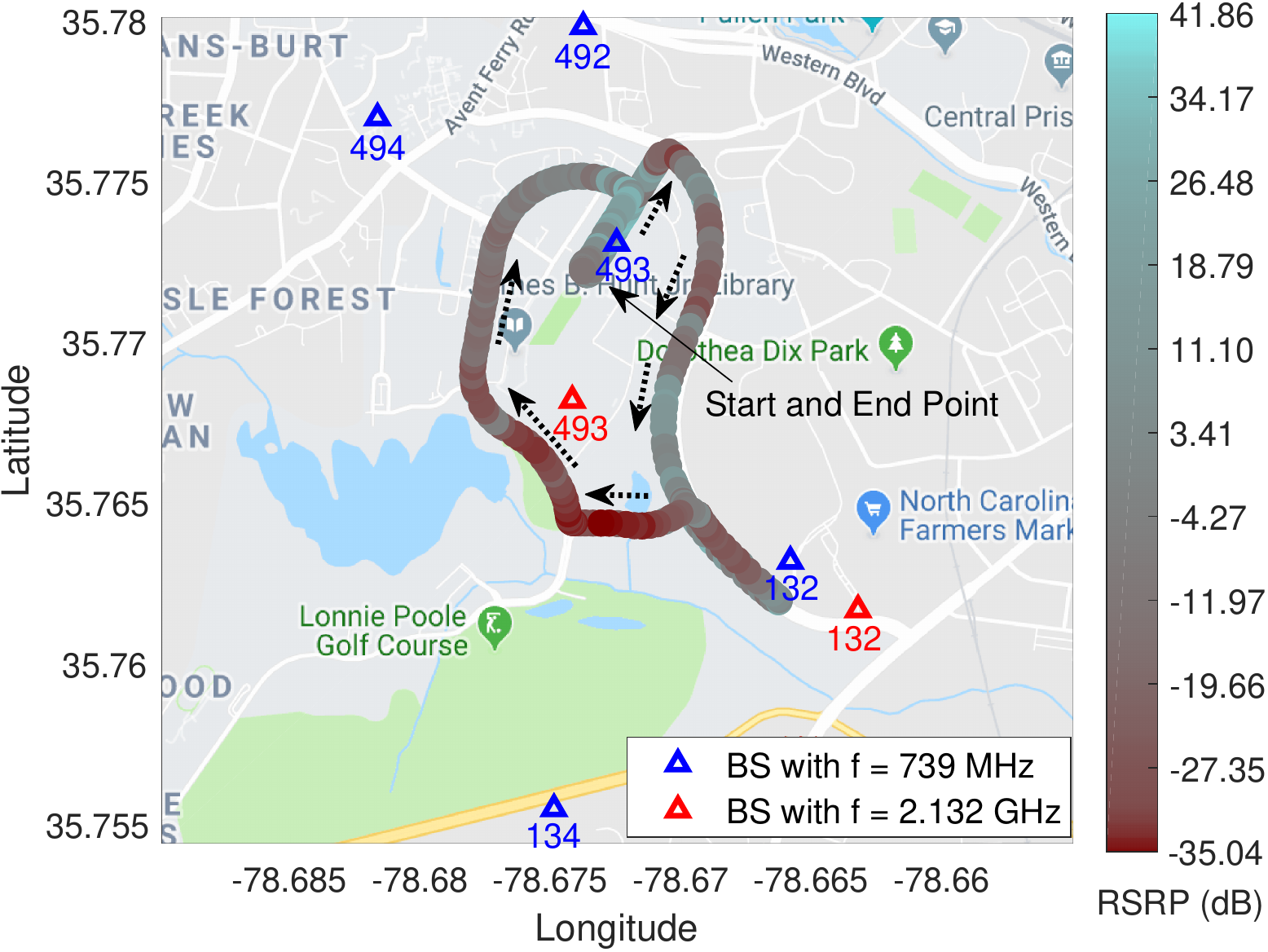}
\label{Fig:Drive_test_NCSU}}\\
\subfloat[{RSRP plots for different BSs in 739~MHz.}]{\includegraphics[width=2.8in]{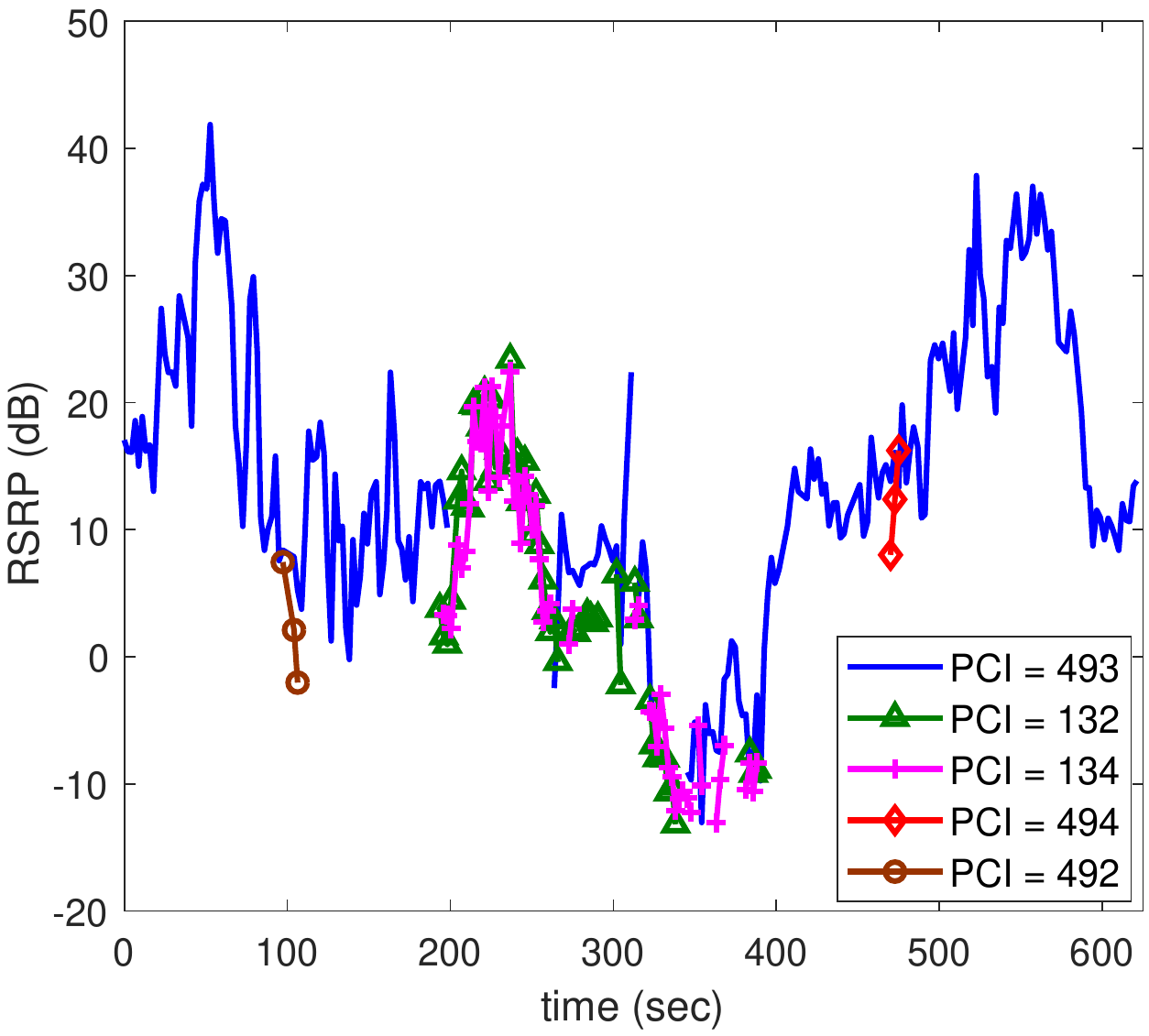}
\label{Fig:RSRP_inplot_NCSU}}\\
\subfloat[{RSRP plots for the strongest BS in different frequencies.}]{\includegraphics[width=2.8in]{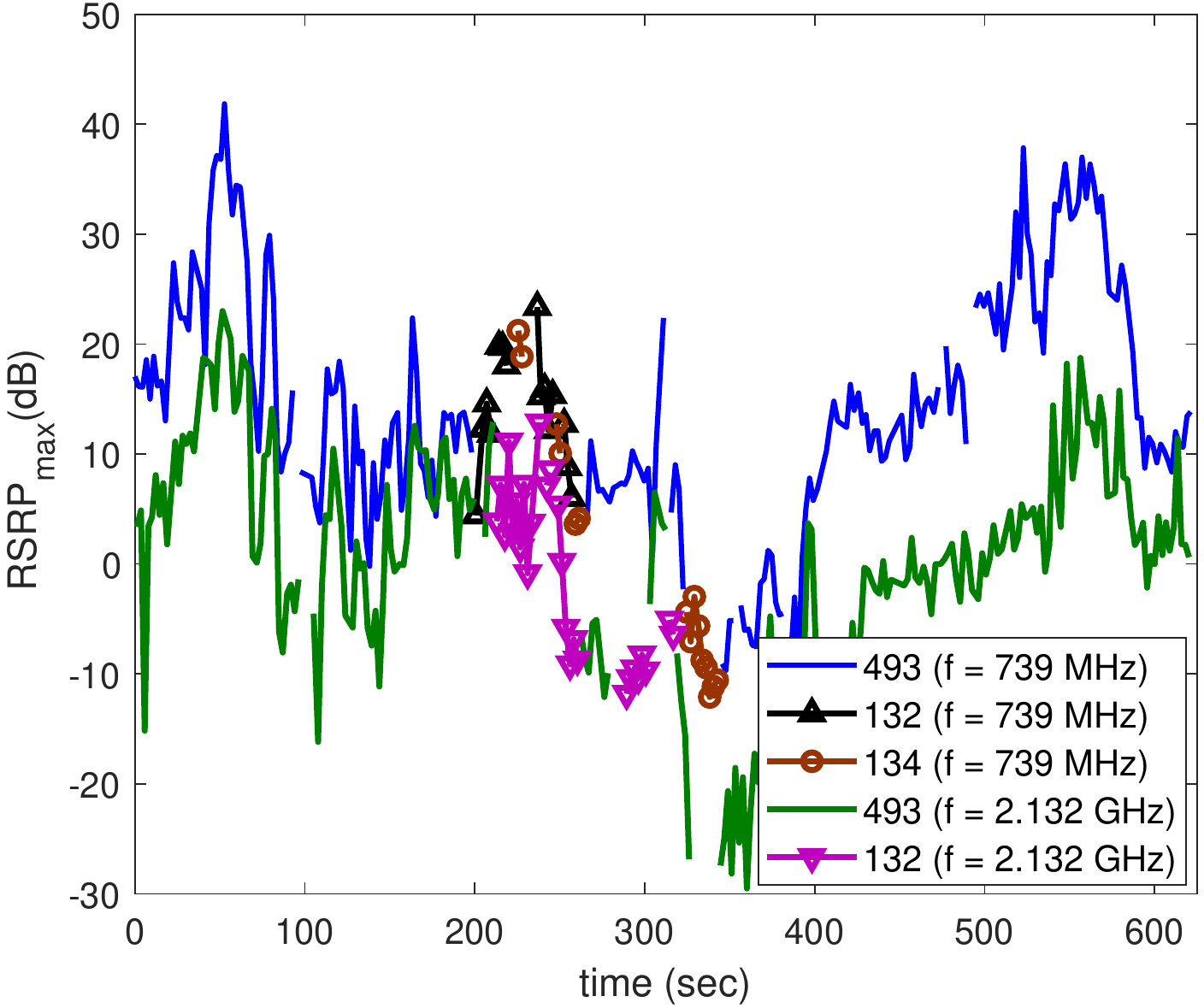}
\label{Fig:RSRP_max_NCSU}}\\
\caption{Drive test results around NCSU campus.}\vspace{-0.2cm}
\label{Fig:Drive_test_results}
\vspace{-3mm}
\end{figure}

The drive test measurements are performed around the North Carolina State University campus, and at each location the USRP carrier frequency is changed alternatively to have one of the two frequencies $f = [739, 2132]$~MHz. The user locations shown in Fig.~\ref{Fig:Drive_test_results}\subref{Fig:Drive_test_NCSU} are plotted using different colors indicating the RSRP value of the strongest decoded BS, while red and blue triangles represent the BS location. First, we investigate the V2I link performance for the frequency measurements at 739~MHz, as shown in Fig.~\ref{Fig:Drive_test_results}\subref{Fig:RSRP_inplot_NCSU}. At the start of the drive test, RSRP for the cell with PCI = 493 is stronger and gradually decreases until the MIB decoding is no longer possible indicating out of coverage. Further, the RSRP values from the cells with PCI = 132/134 appear, and we experience stronger coverage from 493 cell majority of the time compared to rest of the cells. Next, the RSRP value of the strongest BS for the frequencies 739~MHz and 2.132~GHz is shown in Fig.~\ref{Fig:Drive_test_results}\subref{Fig:RSRP_max_NCSU} to compare the V2I link performance. We can see that RSRP values are lower for the frequency 2.132~GHz, and this may be due to increased pathloss for higher frequency or BS might be transmitting with lower power.

\begin{figure}[t]
 \centering
 \includegraphics[width=2.55in]{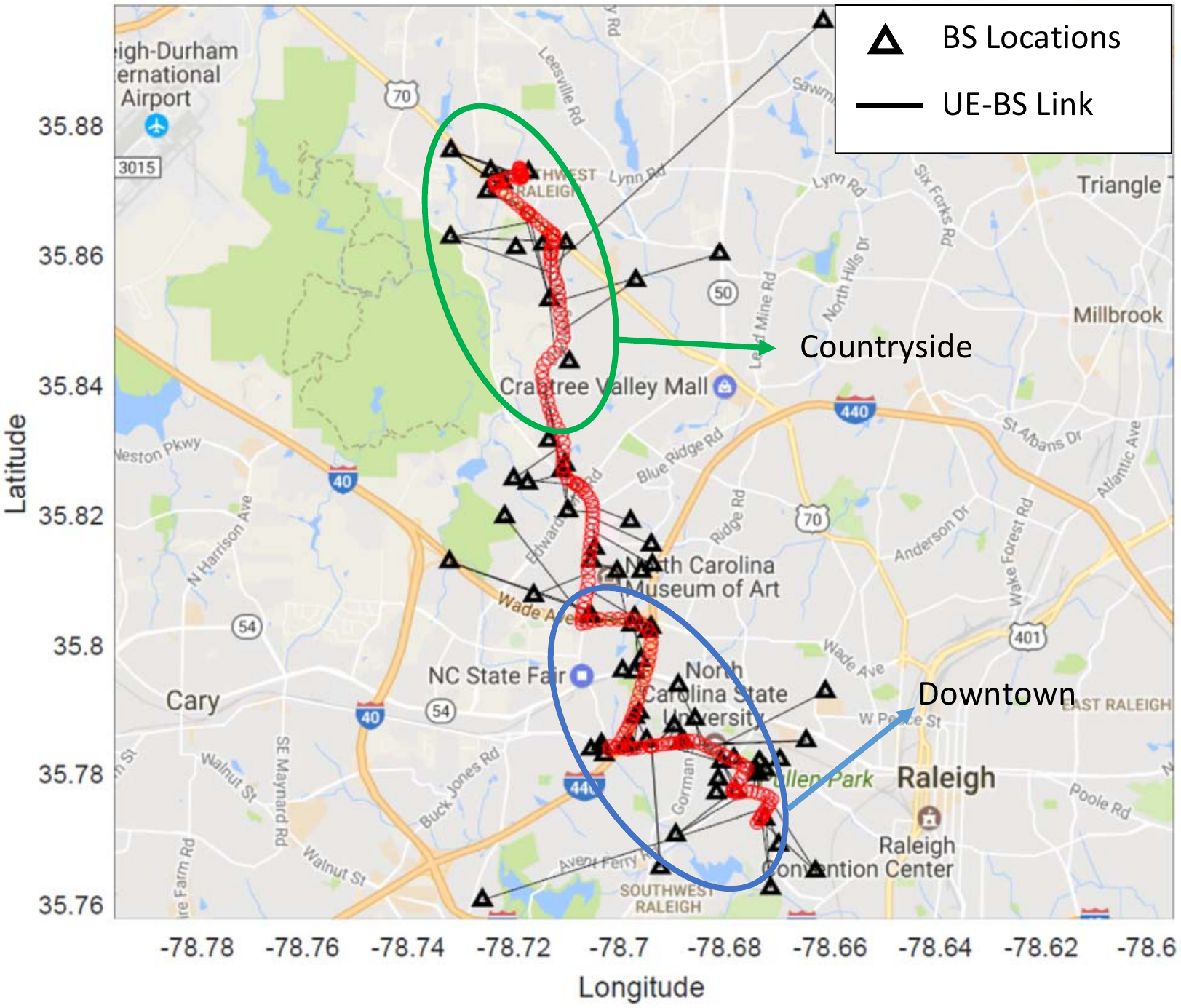}
 \caption{BS distributions during a drive test route in Raleigh, NC.}
 \label{Fig:Drive_test_multfreq_initial}
 \vspace{-3mm}
\end{figure}

Further, the distribution of BSs is examined by obtaining their locations from the measurements performed in the drive test scenario shown in Fig.~\ref{Fig:Drive_test_multfreq_initial}. The USRP carrier frequency is hopped sequentially between frequencies $f = [739, 751, 866.3, 874.4, 882.2, 1967.5, 1992.5, 1982.5]$~MHz at each user location, and the obtained BS locations from these frequency measurements are shown as black triangles. We observe more BS deployments near the Raleigh downtown compared to countryside areas.

\vspace{-3mm}
\subsection{V2I Connectivity Analysis in Different LTE Deployment Scenarios} \label{Subsec:Large_area_analysis}

\begin{figure}[h]
 \centering
 \includegraphics[width=2.83in]{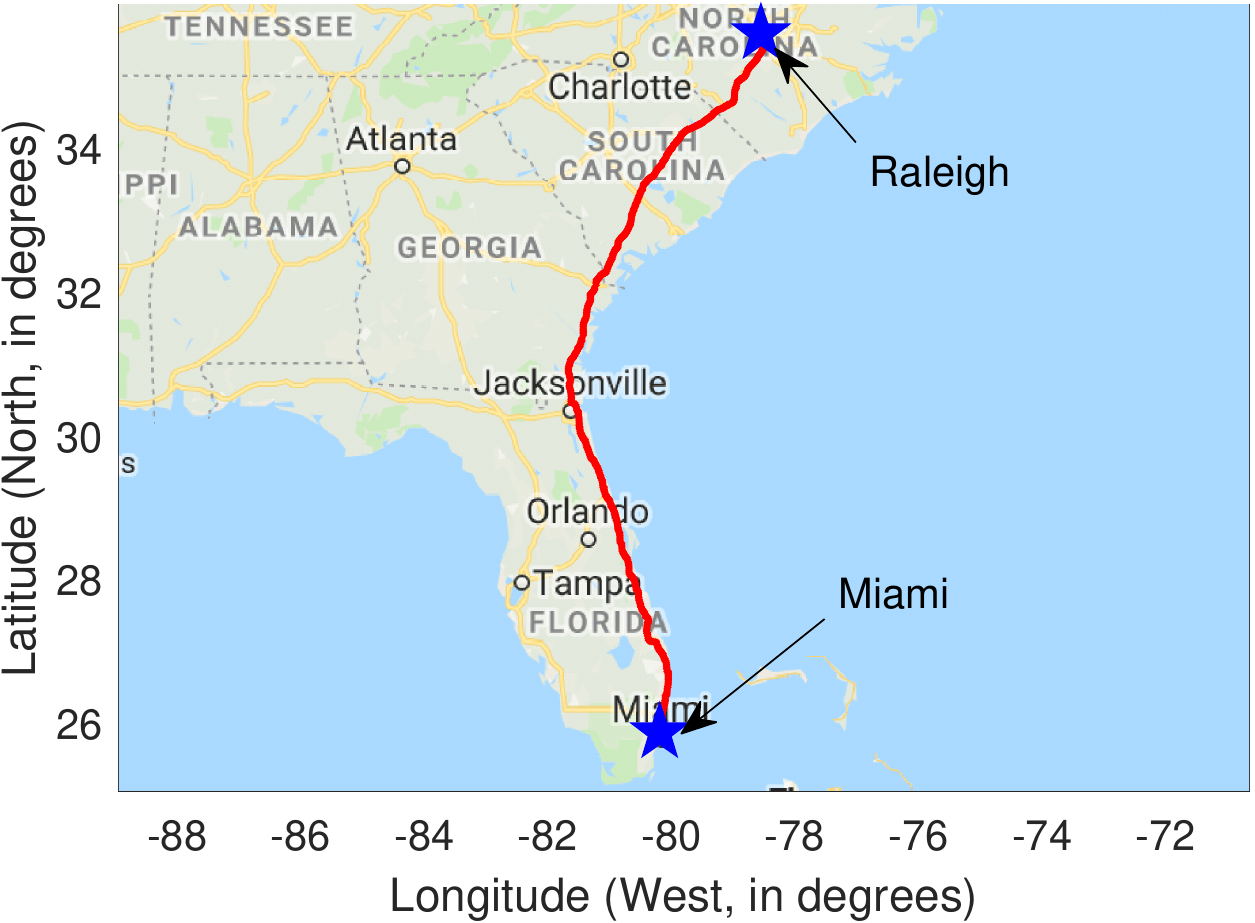}
 \caption{Drive test route between Raleigh, NC and Miami, FL, covering urban, sub-urban, and rural environments.}
 \label{Fig:drive_test_route}
\end{figure}

The drive test measurement campaign is conducted from Miami to Raleigh covering four states, namely Florida, Georgia, South Carolina, and North Carolina as shown in Fig.~\ref{Fig:drive_test_route}.
The central frequency of the measurements were fixed at 739~MHz, and the results are acquired on a county level, which are then categorized in to urban, semi-urban and rural types based on population census.

\begin{figure}[t]
    \centering
\subfloat[{RSRP distribution of different counties.}]{\includegraphics[width=2.76in]{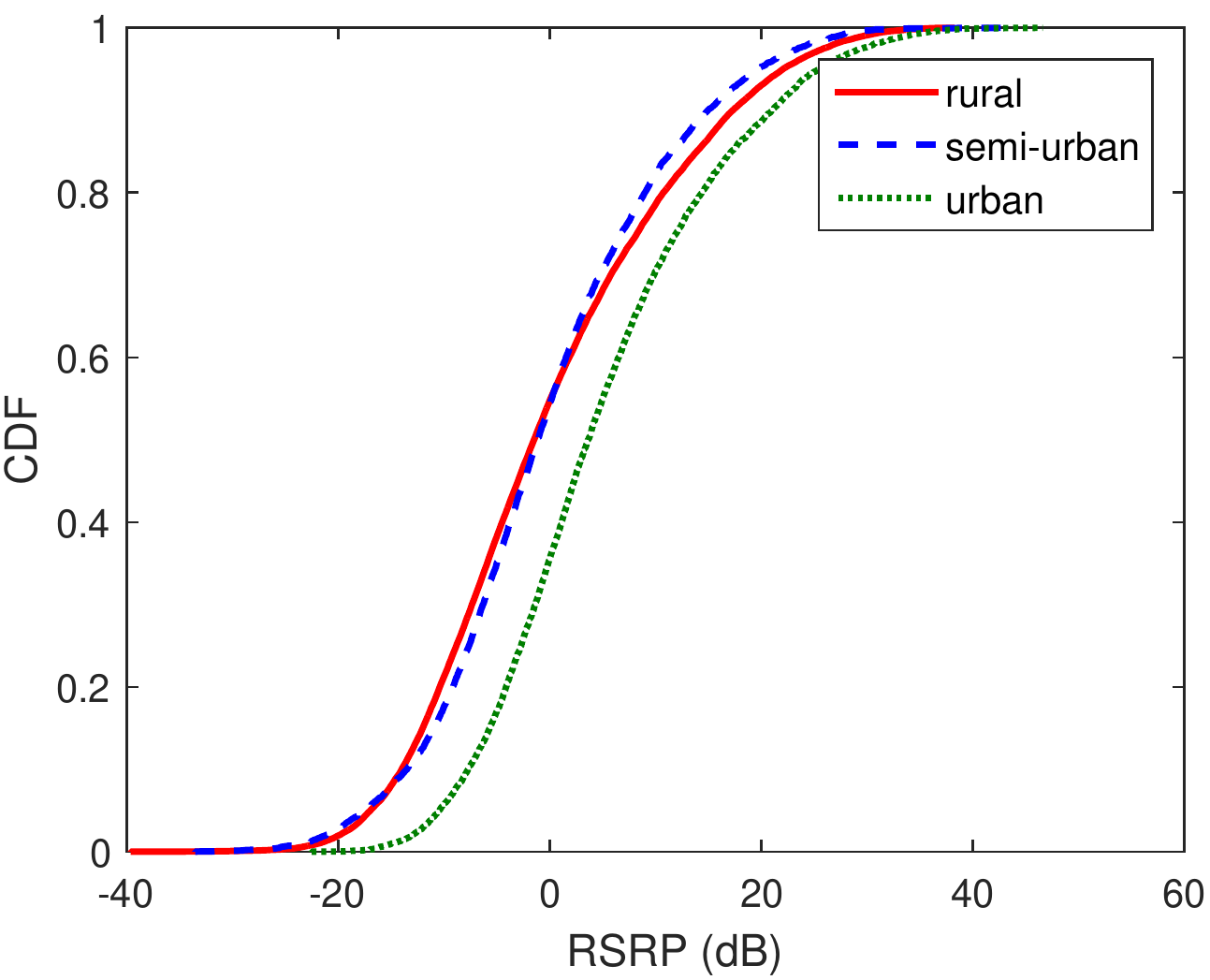}
\label{Fig:RSRP_CDF}}\\
\subfloat[{BS to user distance distribution of different counties.}]{\includegraphics[width=2.76in]{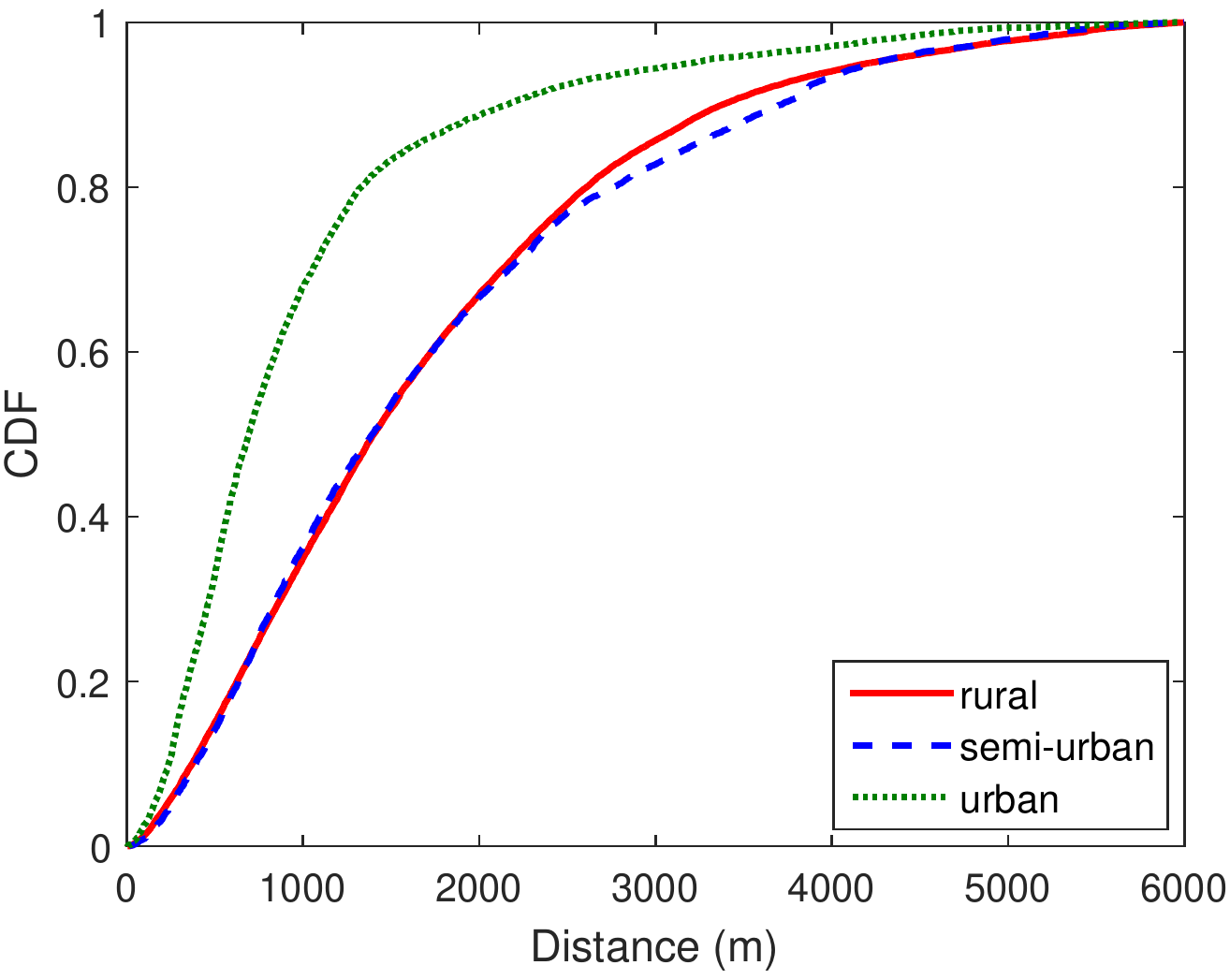}
\label{Fig:Distance_CDF}}\\
\subfloat[{Vehicle velocity distribution in different counties.}]{\includegraphics[width=2.76in]{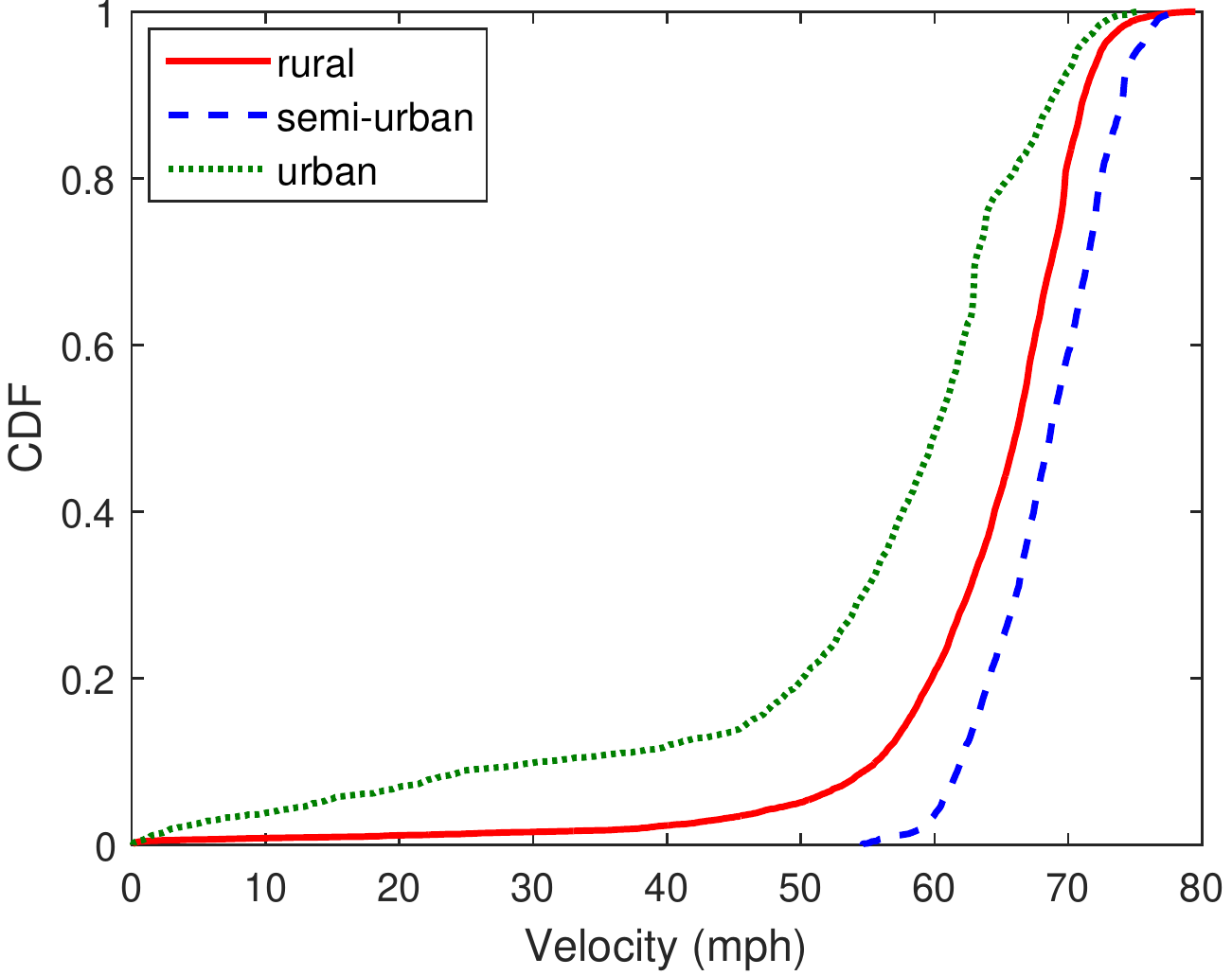}
\label{Fig:Velocity_CDF}}
\caption{Drive test scenario results in different counties.}
    \label{Fig:Drive_large_scale_results}
    \vspace{-4mm}
\end{figure}

First, cumulative distribution of RSRP results is shown in Fig.~\ref{Fig:Drive_large_scale_results}\subref{Fig:RSRP_CDF} to investigate the V2I connectivity performance of LTE networks. We observe that the RSRP values tend to be greater in urban areas indicating stronger V2I connectivity compared to semi urban and rural counties. This is because of the shorter distances between a BS and a user which is verified from the cumulative distribution plots shown in Fig.~\ref{Fig:Drive_large_scale_results}\subref{Fig:Distance_CDF}. Next, the velocity statistics of all the counties is examined from the cumulative distribution shown in Fig.~\ref{Fig:Drive_large_scale_results}\subref{Fig:Velocity_CDF}. We observe higher possibility of larger velocities in semi-urban compared to rural and urban counties. This is because of lower semi-urban counties in the drive test campaign and therefore sparse result data are obtained for semi-urban counties.



\begin{figure}[t]
    \centering
    \subfloat[{Joint distribution of RSRP and BS distance in rural counties.}]{\includegraphics[trim=0 0 0.3cm 0, clip,width=2.9in]{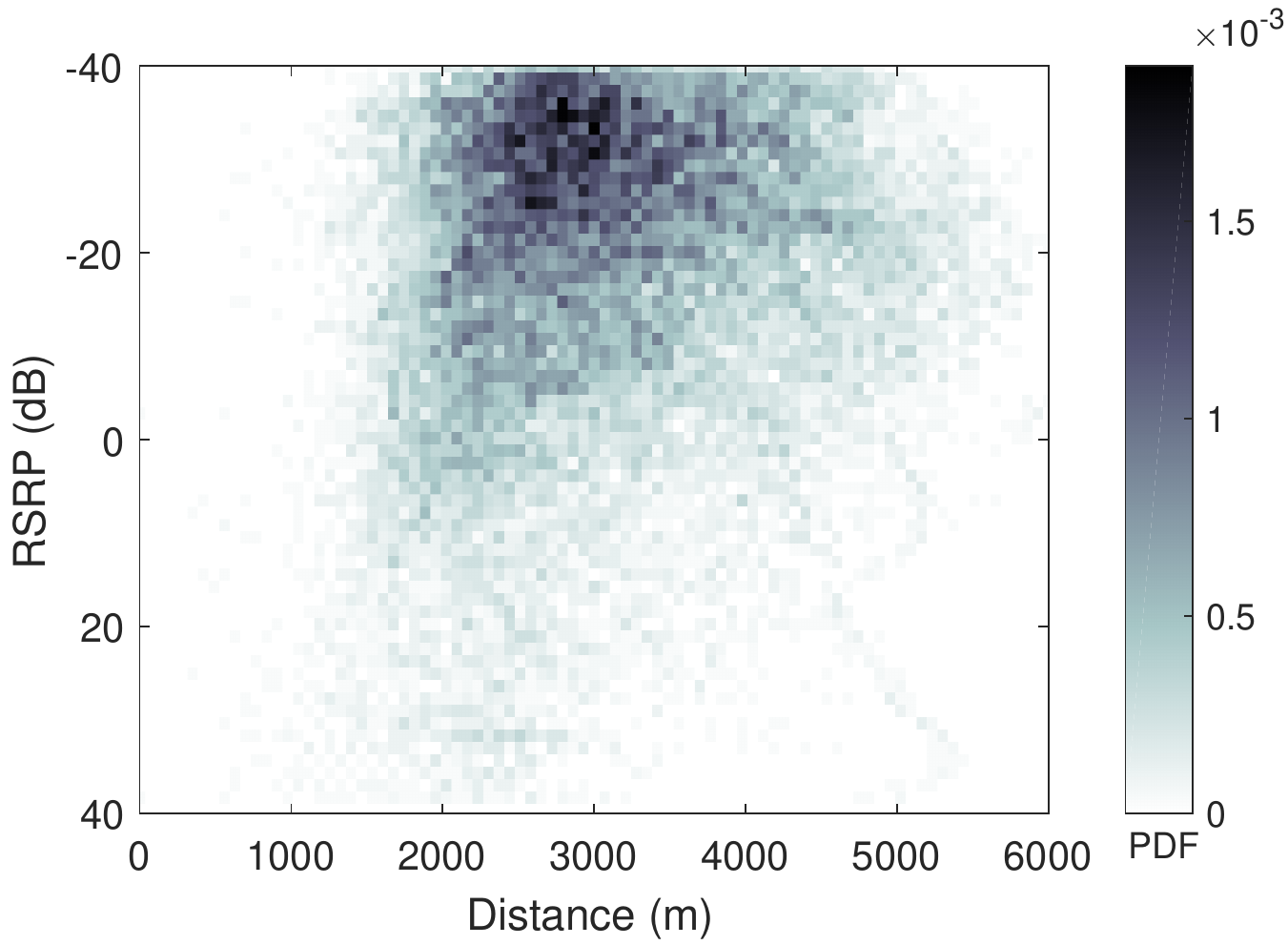}
\label{Fig:RSRP_dist_rur}} \\
\subfloat[{Joint distribution of RSRP \& BS distance in semi-urban counties.}]{\includegraphics[trim=0 0 0.3cm 0, clip,width=2.9in]{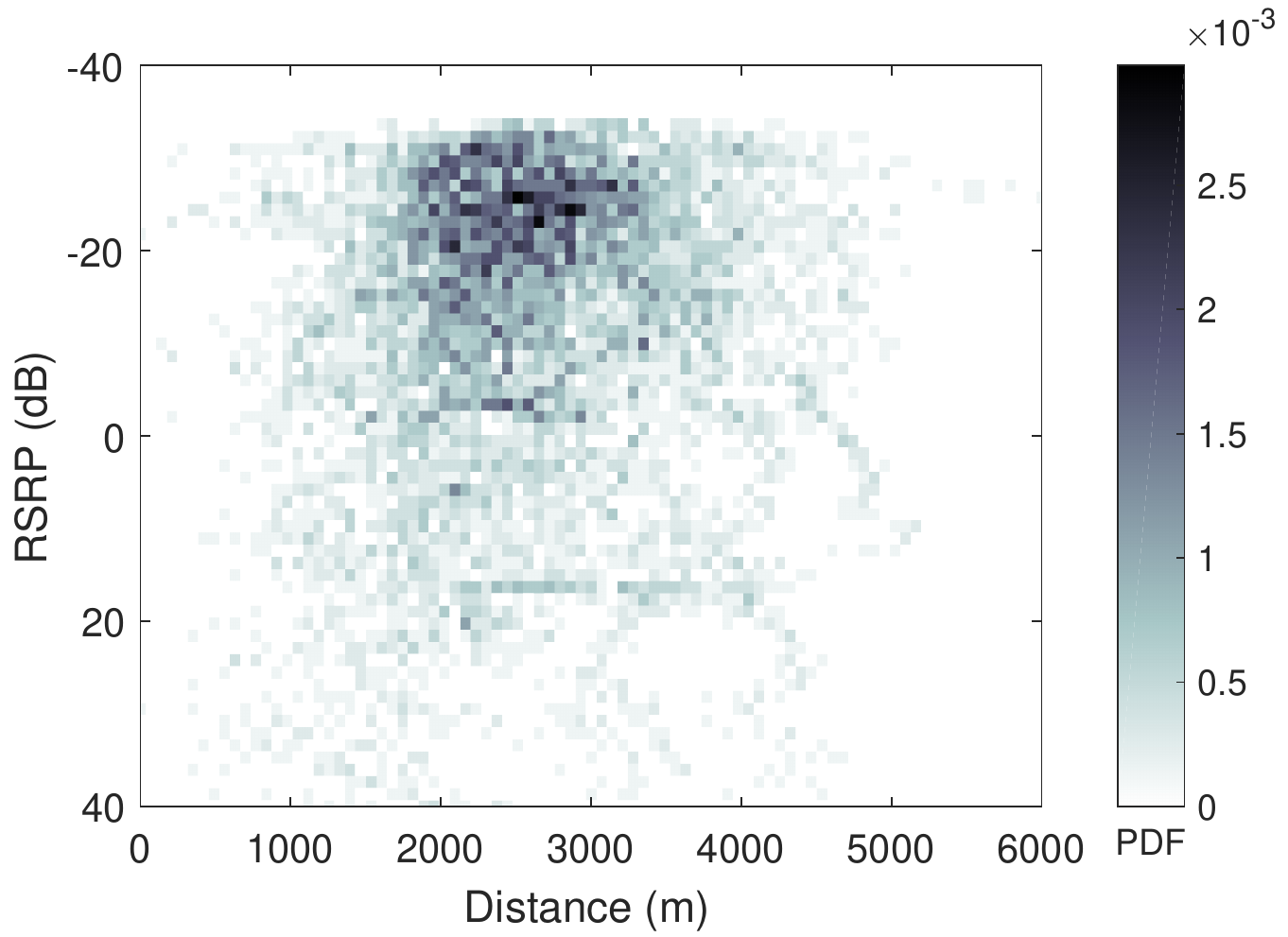}
\label{Fig:RSRP_dist_semi}}\\
\subfloat[{Joint distribution of RSRP and BS distance in urban counties.}]{\includegraphics[trim=0 0 0.3cm 0, clip,width=2.9in]{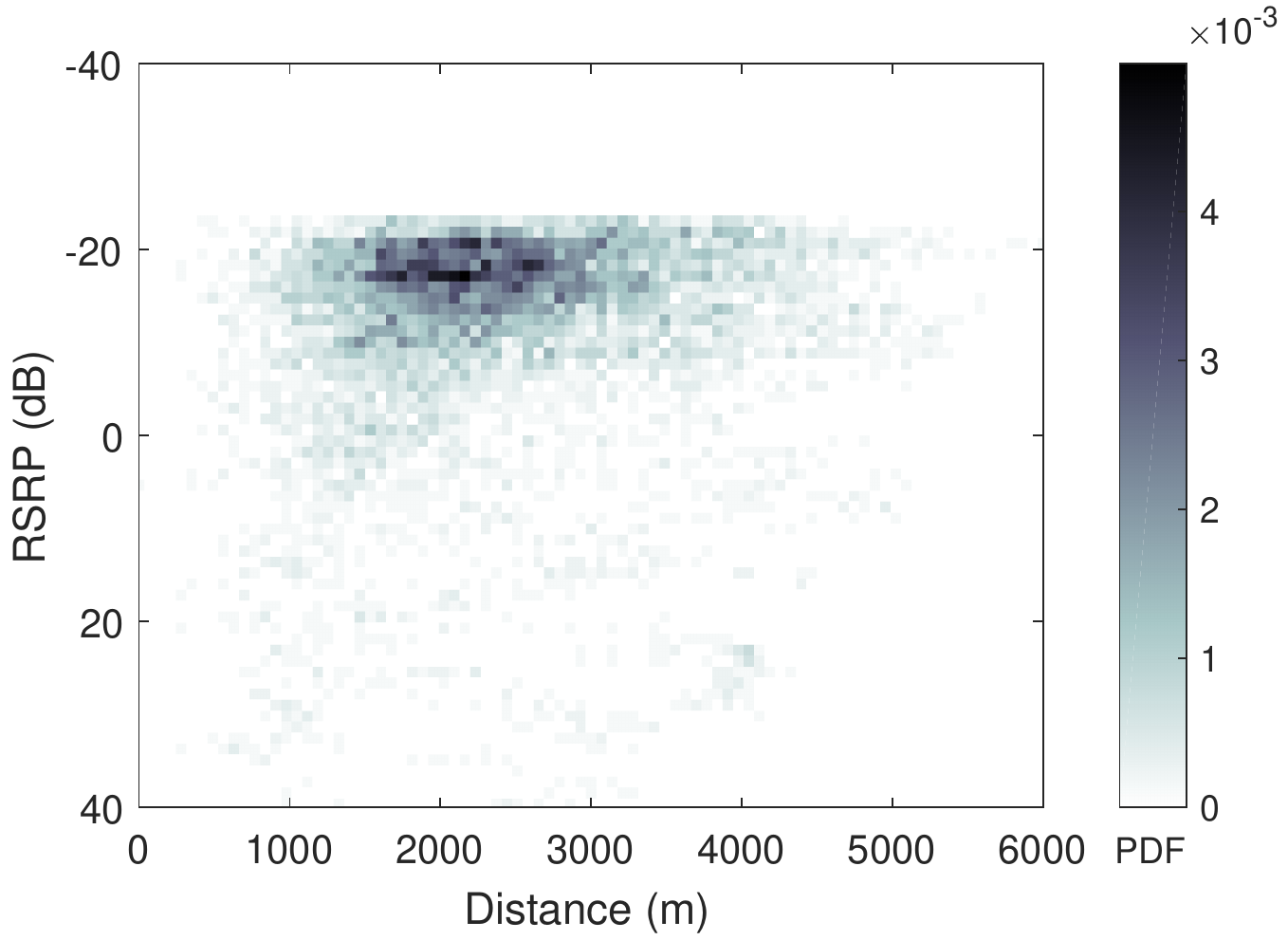}
\label{Fig:RSRP_dist_urb}}
\caption{Joint distribution of RSRP and BS-distance in different counties.}
    \label{fig:RSRP_dist_pdf}
    \vspace{-4mm}
\end{figure}

Further, joint distribution of RSRP and BS distance plots for urban, semi-urban and rural counties are shown in Fig.~\ref{fig:RSRP_dist_pdf}. We observe a greater likelihood of BS distance around 3000~m with RSRP varying from -40 dB to 40 dB in rural counties. On the other hand, it is more likely to have a smaller distance between a BS and a user in urban counties, and it is around 2000~m. Also stronger RSRP values are observed varying from -20 dB to 40 dB, in urban counties.




\begin{figure}[t]
    \centering
    \subfloat[Joint distribution of velocity and RSRP in rural counties.]{\includegraphics[width=3in]{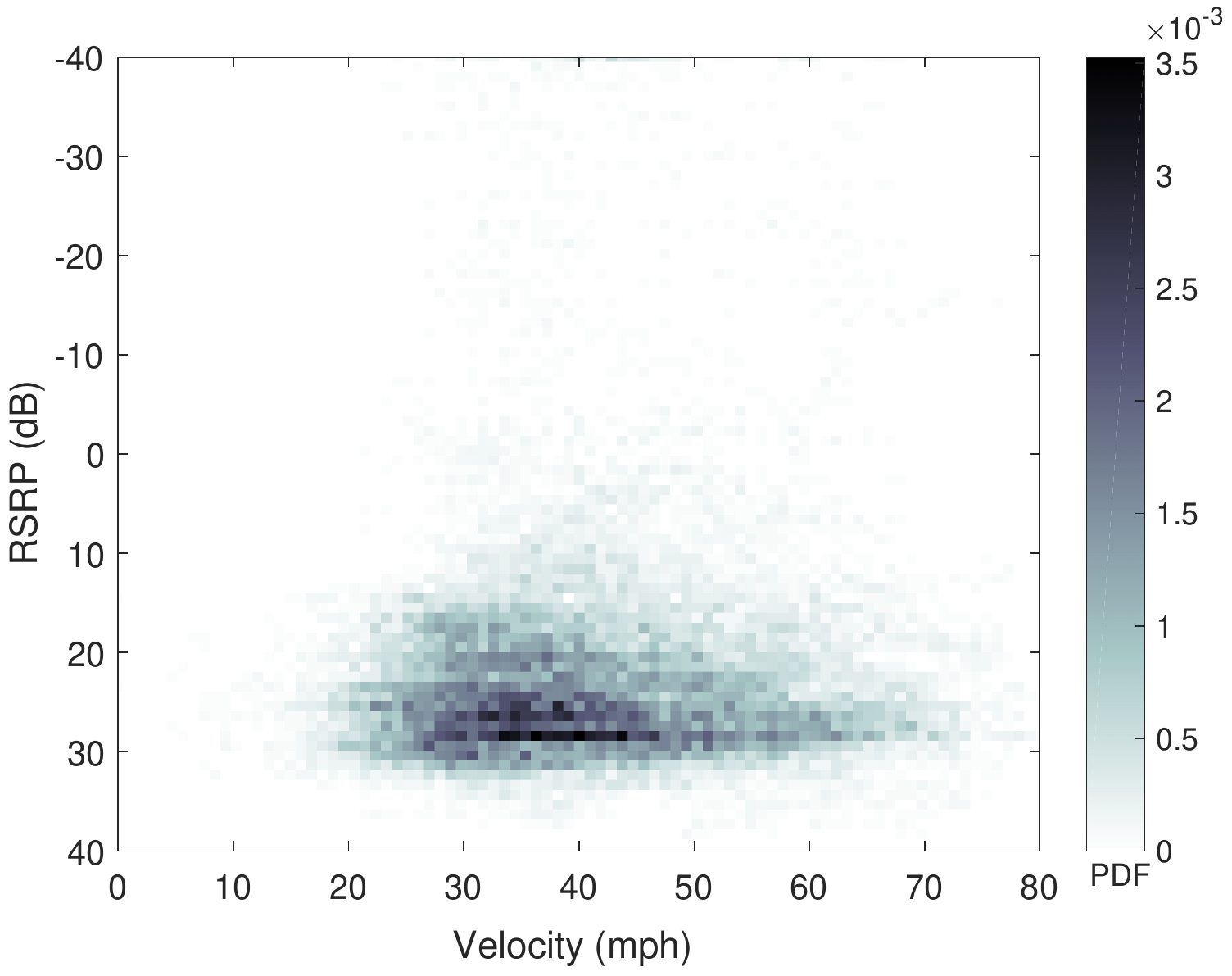}
\label{Fig:vel_RSRP_rur}}\\
\subfloat[Joint distribution of velocity and RSRP in semi-urban counties.]{\includegraphics[width=3in]{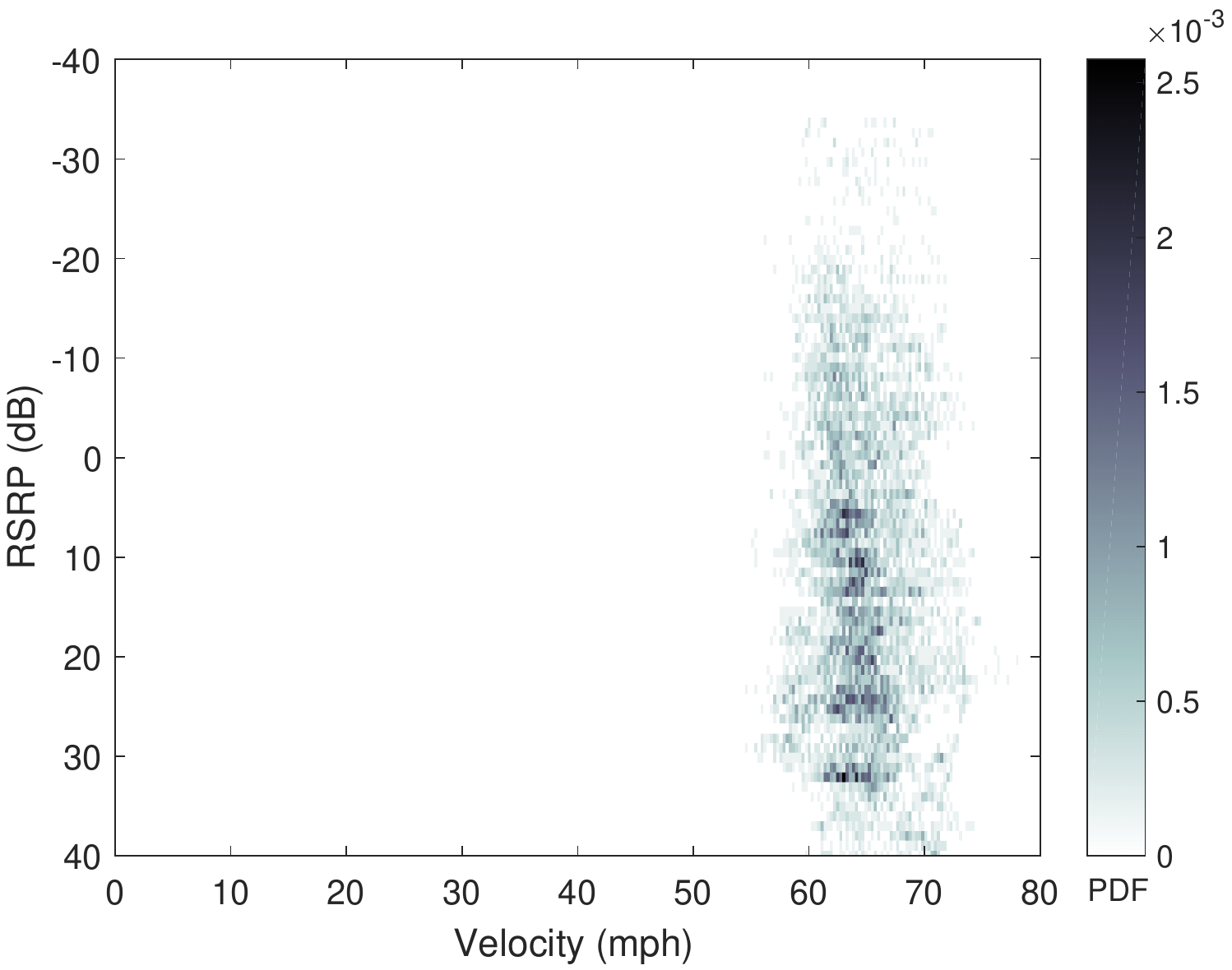}
\label{Fig:vel_RSRP_semi}}\\
\subfloat[Joint distribution of velocity and RSRP in urban counties.]{\includegraphics[width=3in]{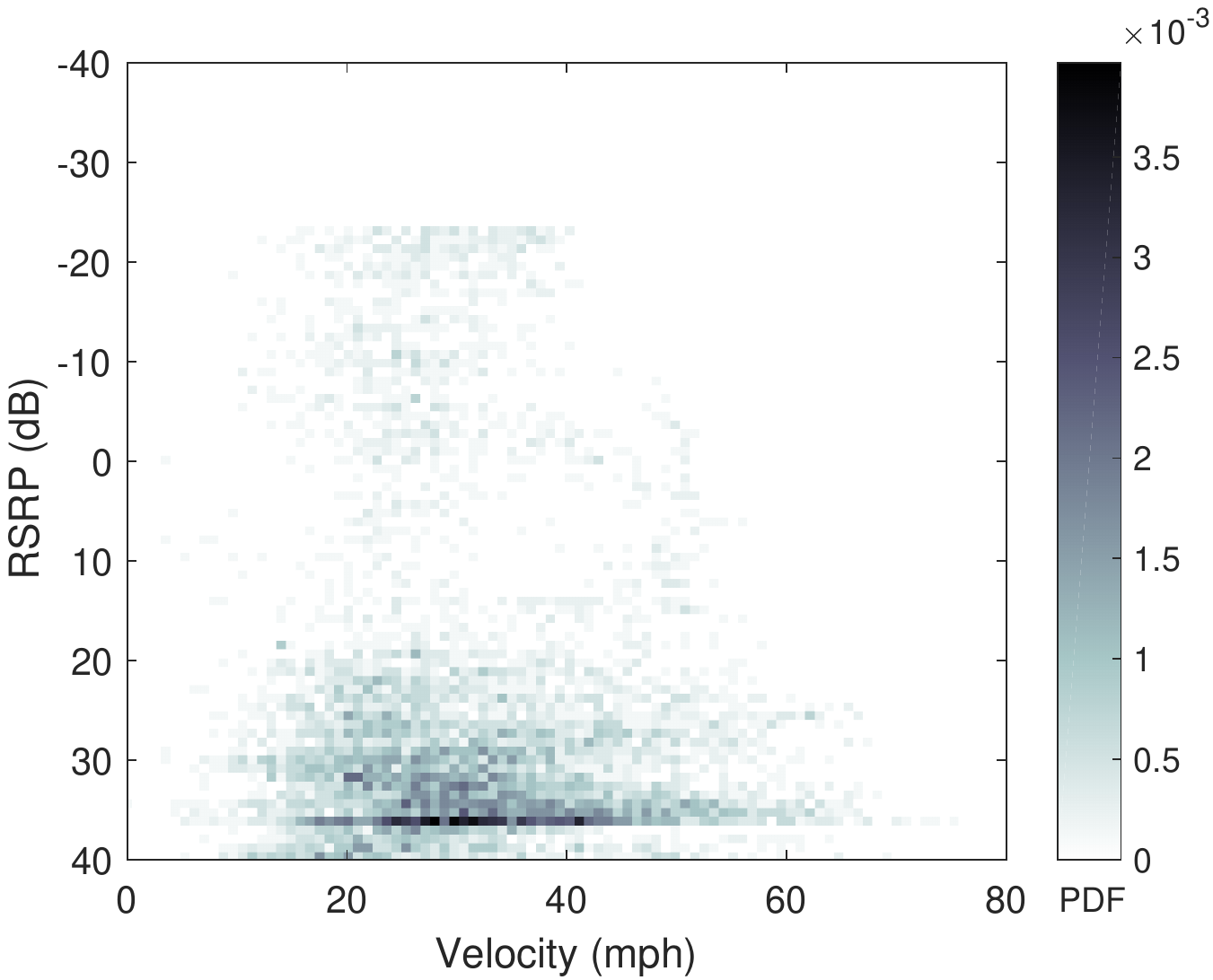}
\label{Fig:vel_RSRP_urb}}
    \caption{Joint RSRP-velocity distribution in different counties.}
    \label{fig:vel_RSRP_pdf}
    \vspace{-5mm}
\end{figure}

Further, impact of the user velocity on V2I link performance is studied from the joint distribution of RSRP, and results are shown in Fig.~\ref{fig:vel_RSRP_pdf}. We observe greater likelihood of lower velocities in rural and urban compared to semi-urban counties. In addition, RSRP tends to fluctuate with wider range for user velocities in semi-urban and rural compared to urban counties.




\begin{figure}[t]
    \centering
    \subfloat[Joint distribution of velocity and BS distance in rural counties]{\includegraphics[width=3.2in]{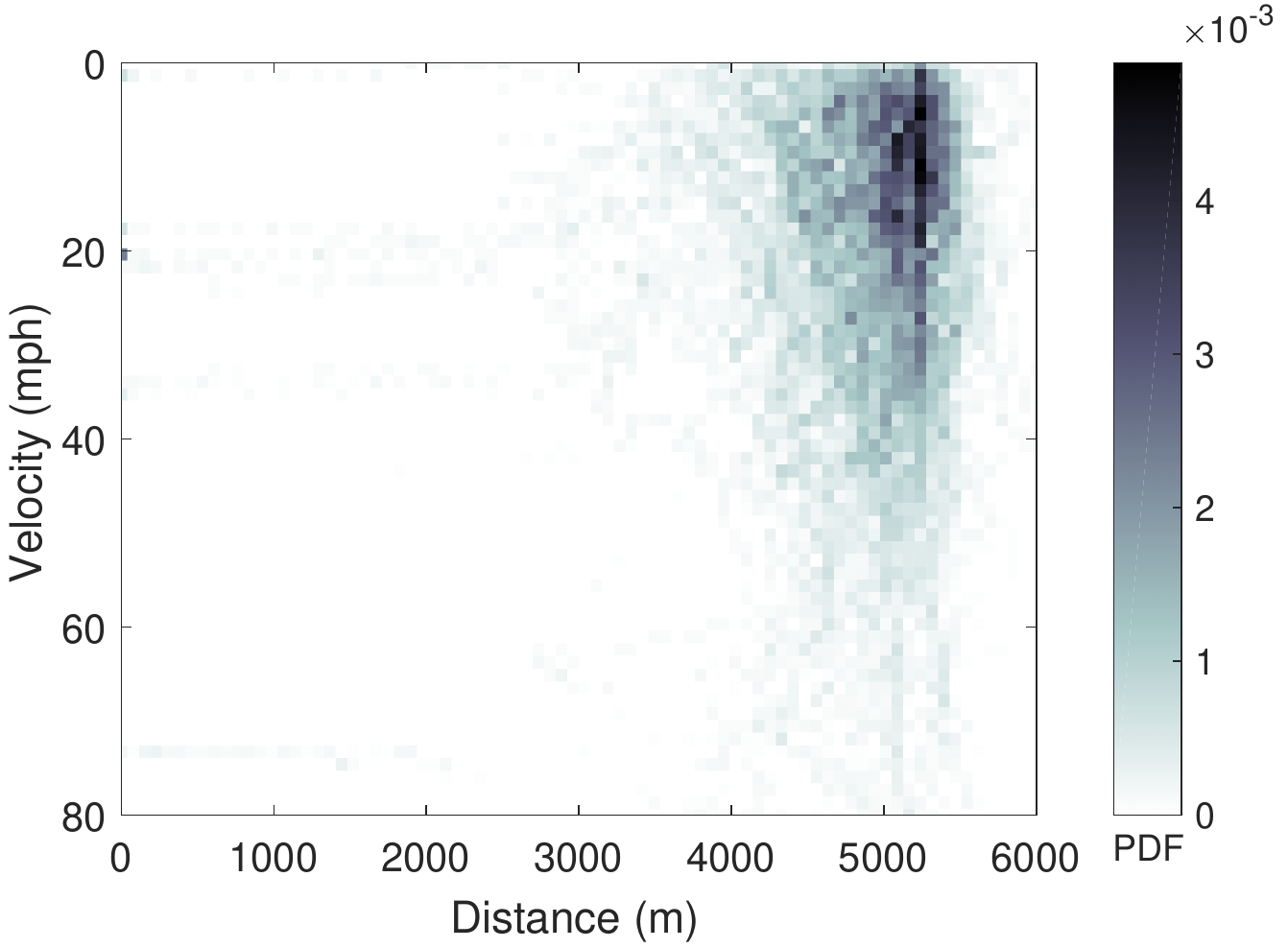}
\label{Fig:vel_dist_rur}}\\
\subfloat[Joint distribution of velocity and BS distance in semi-urban counties]{\includegraphics[width=3.2in]{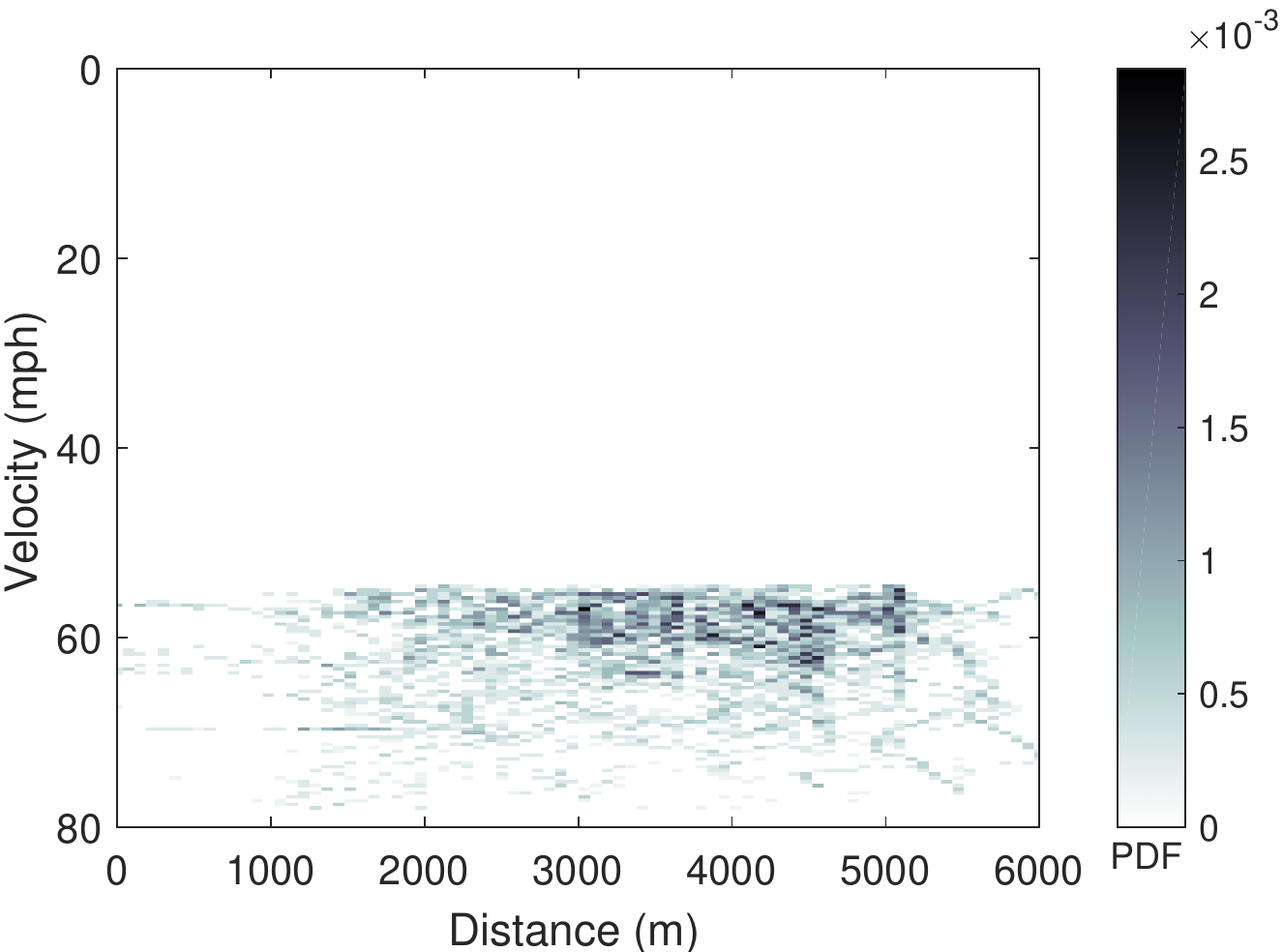}
\label{Fig:vel_dist_semi}}\\
\subfloat[Joint distribution of velocity and BS distance in urban counties]{\includegraphics[width=3.2in]{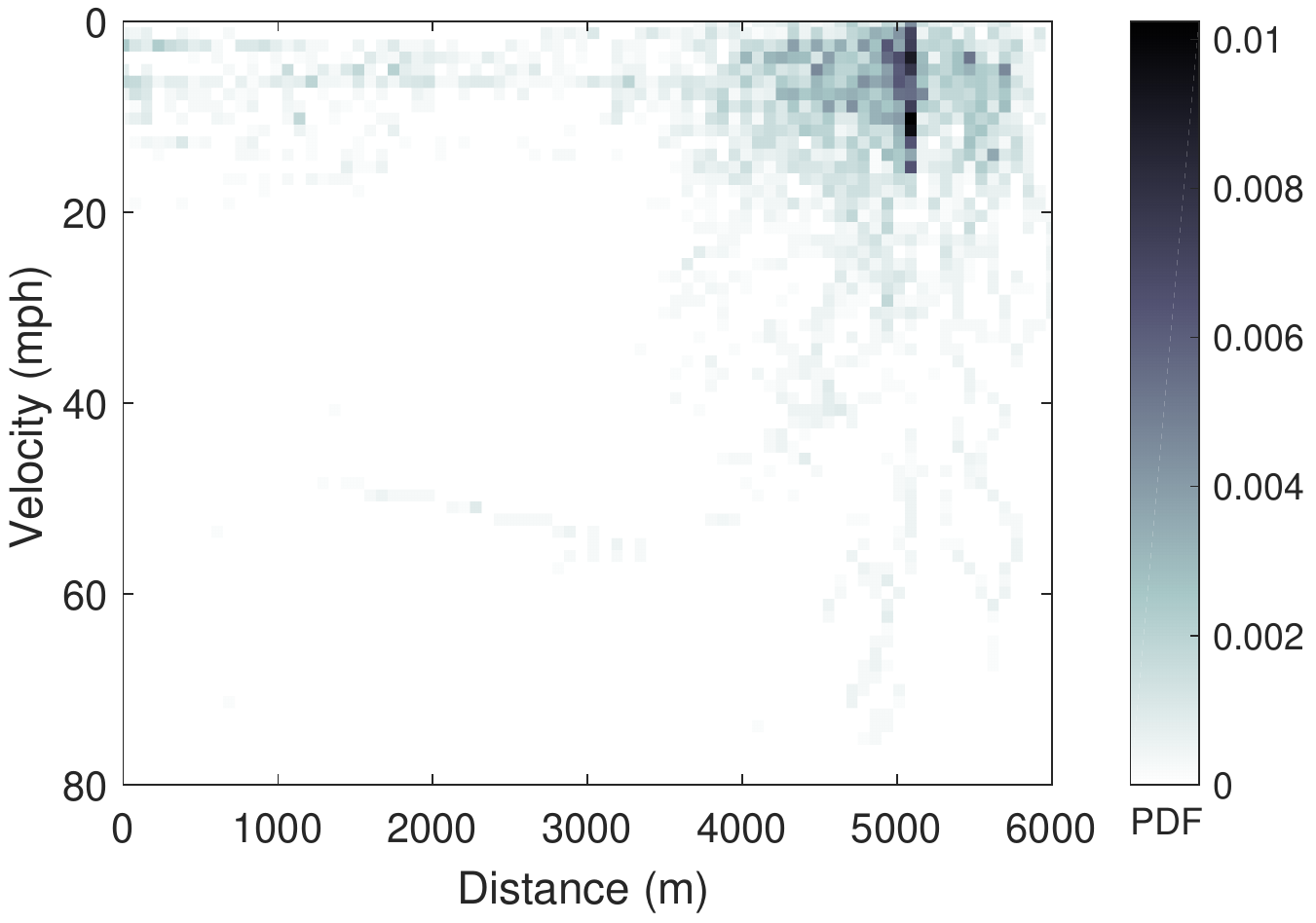}
\label{Fig:vel_dist_urb}}
    \caption{Velocity distance distribution in different counties.}
    \label{fig:vel_dist_pdf}
    \vspace{-4mm}
\end{figure}

Lastly, the joint distribution of velocity and BS distance is shown in Fig.~\ref{fig:vel_dist_pdf}. We observe a higher likelihood of BS distance close to 5000~m in urban counties while in rural counties the chance of BS distances are greater than 5000~m.



\begin{figure}[t]
    \centering
    \subfloat[Disconnected duration profile in rural counties]{\includegraphics[width=2.89in]{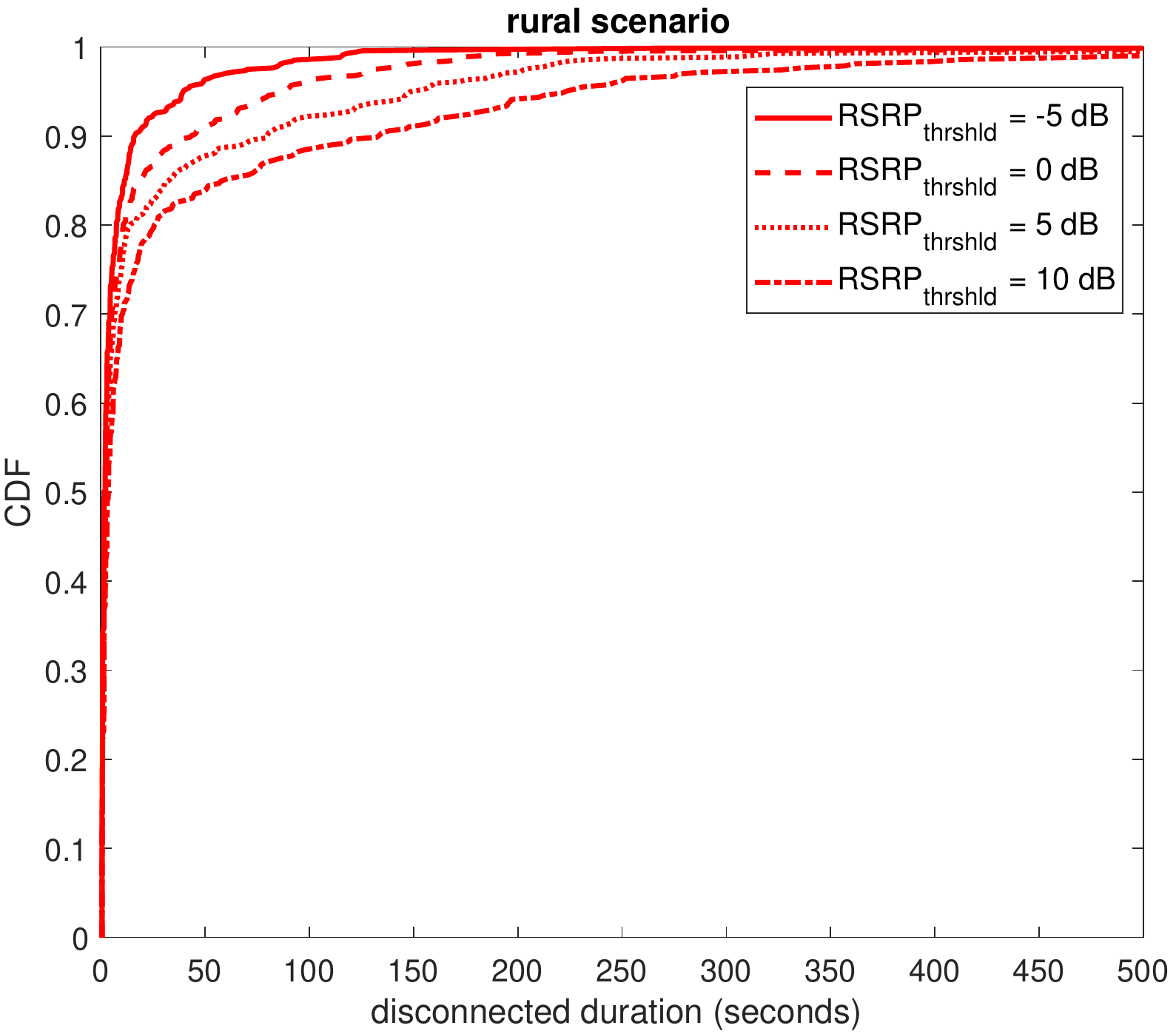}
\label{Fig:dis_dur_rur}}\\
\vspace{-1mm}
\subfloat[Disconnected duration profile in semi-urban counties]{\includegraphics[width=2.89in]{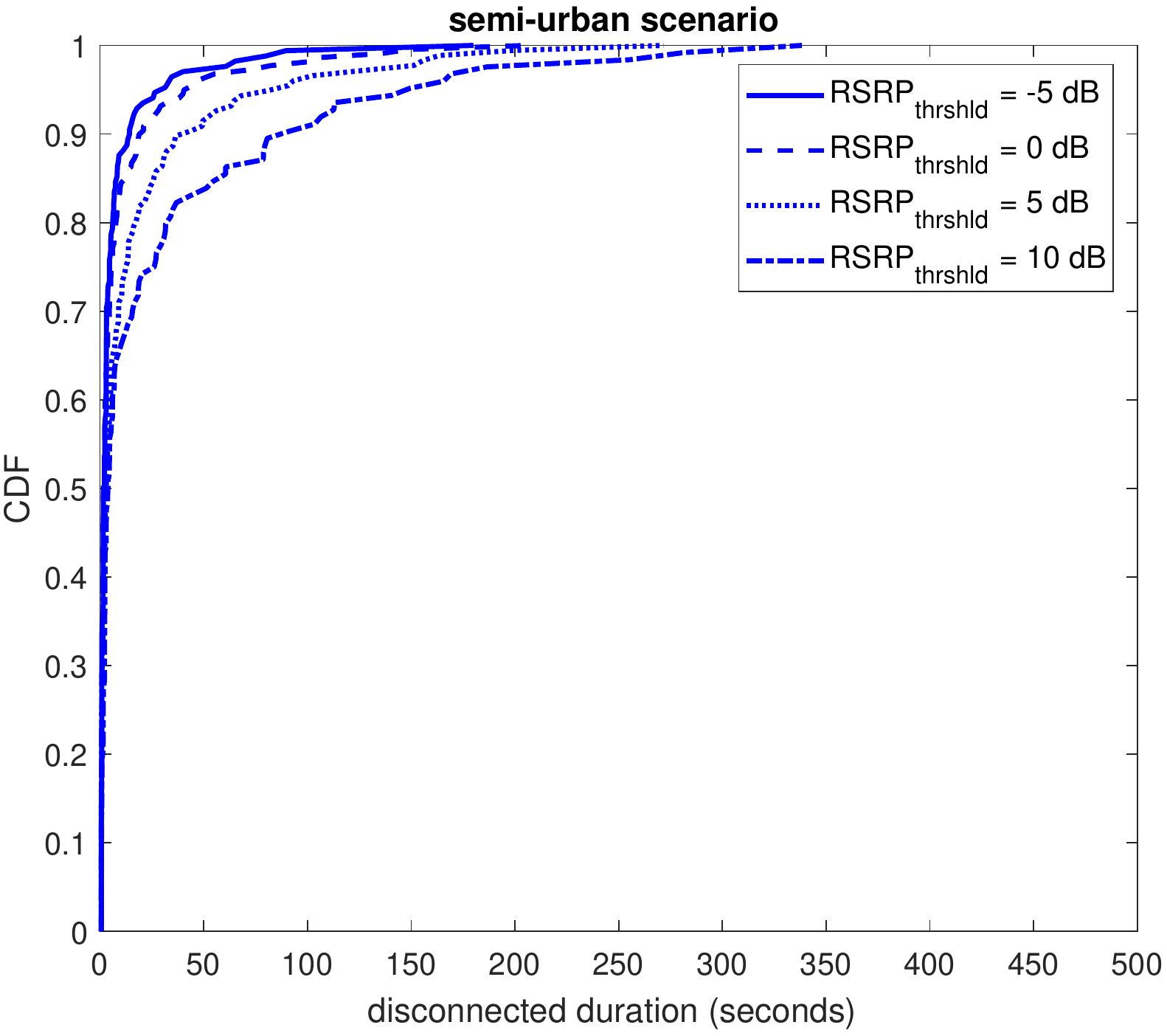}
\label{Fig:dis_dur_semi}}\\
\vspace{-1mm}
\subfloat[Disconnected duration profile in urban counties]{\includegraphics[width=2.89in]{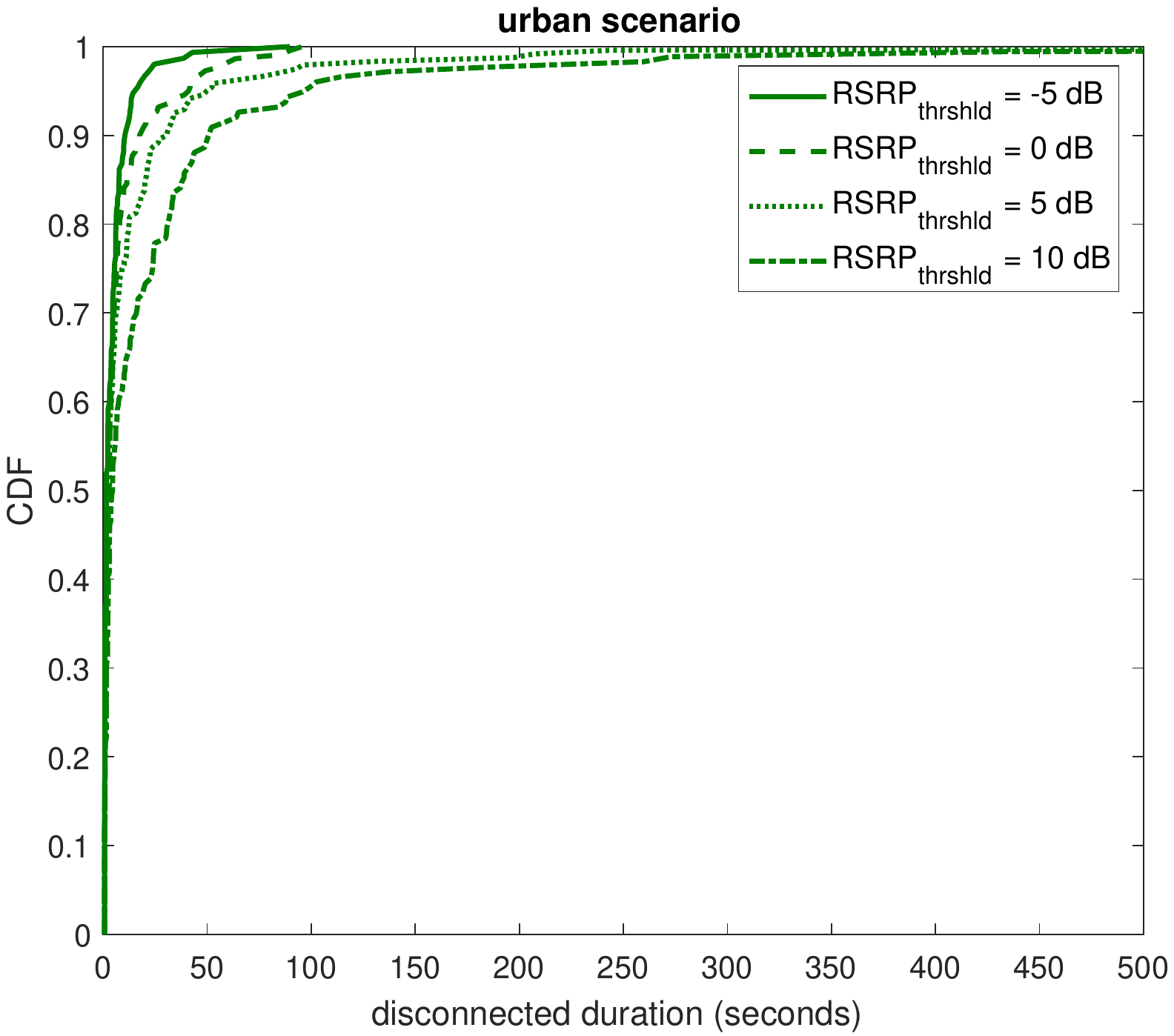}
\label{Fig:dis_dur_urb}}
    \caption{Disconnected duration distributions in different environments.}
    \label{fig:dis_dur_cdf}
    \vspace{-6mm}
\end{figure}

Next, the time duration statistics of RSRP from BSs is extracted to study the disconnected duration profile. It is the duration where we do not observe stronger RSRP values from any BS from the fixed threshold. The disconnected duration is evaluated considering different RSRP thresholds and its plots for rural, semi-urban and urban counties are shown in Fig.~\ref{fig:dis_dur_cdf}. In general, we observe higher chance of disconnected duration with increase in RSRP threshold. Additionally, we see possibility of disconnected duration in urban counties is lower compared to rural and semi-urban~counties.

\section{Conclusion}
In this paper, we examined the V2I connectivity performance of LTE deployments from the extensive drive test campaign measurements. Mainly we assessed the coverage performance of LTE networks in rural, semi-urban and urban scenarios. The results show stronger coverage and shorter duration of disconnectivity in urban and semi-urban compared to rural scenarios. In addition, the higher operating frequencies of LTE networks weaken the V2I connectivity. We provide our developed SDR codes for other researchers' use in~\cite{Matlab_LV_code}. 


In future work, the drive test campaign results will be incorporated into an LTE system level simulator to assess the mobility performance of LTE networks~\cite{vasu_TVT_2014,vasufuzzy_IEEEA_2017,6215543}. We plan to evaluate key mobility performance indicators and investigate the impact of channel congestion on V2I connectivity in road safety applications.

\section{Acknowledgment}

The authors would like to thank Nadisanka Rupasinghe for his help in developing SDR codes. This work was supported in part by the National Science Foundation under
the award numbers CNS-1453678 and CNS-1814727.


\bibliography{Ref_Exp_Mobility_LTE}
\bibliographystyle{IEEEtran}









\end{document}